\documentclass[letterpaper]{article} 
\usepackage{aaai2026}  
\usepackage{times}  
\usepackage{helvet}  
\usepackage{courier}  
\usepackage[hyphens]{url}  
\usepackage{graphicx} 
\usepackage{natbib}  
\usepackage{caption} 
\usepackage{algorithm}
\usepackage{algorithmic}
\usepackage{amsmath} 

\usepackage{amssymb} 
\usepackage{amsfonts} 
\usepackage{bm}
\usepackage{booktabs}
\usepackage{amsmath}
\usepackage{tabularx} 
\usepackage{array}
\usepackage[table]{xcolor} 
\usepackage{makecell}

\usepackage{newfloat}
\usepackage{listings}
\DeclareCaptionStyle{ruled}{labelfont=normalfont,labelsep=colon,strut=off} 
\floatstyle{ruled}
\newfloat{listing}{tb}{lst}{}
\floatname{listing}{Listing}
\title{Unsupervised Single-Channel Audio Separation with Diffusion Source Priors}
\author{
    Runwu Shi\textsuperscript{\rm 1}, Chang Li\textsuperscript{\rm 2}, Jiang Wang\textsuperscript{\rm 1}, Rui Zhang\textsuperscript{\rm 3}, Nabeela Khan\textsuperscript{\rm 1}, \\
    Benjamin Yen\textsuperscript{\rm 1}, Takeshi Ashizawa\textsuperscript{\rm 1}, Kazuhiro Nakadai\textsuperscript{\rm 1}
}
\affiliations{
    \textsuperscript{\rm 1}Department of Systems and Control Engineering, Institute of Science Tokyo
    \\
    \textsuperscript{\rm 2}University of Science and Technology of China
    \\
    \textsuperscript{\rm 3}University of Hong Kong
    \\
    \{shirunwu, nakadai\}@ra.sc.e.titech.ac.jp
}

\usepackage{bibentry}

\begin{document}

\maketitle

\begin{abstract}
Single-channel audio separation aims to separate individual sources from a single-channel mixture. Most existing methods rely on supervised learning with synthetically generated paired data. However, obtaining high-quality paired data in real-world scenarios is often difficult. This data scarcity can degrade model performance under unseen conditions and limit generalization ability. To this end, in this work, we approach this problem from an unsupervised perspective, framing it as a probabilistic inverse problem. Our method requires only diffusion priors trained on individual sources. Separation is then achieved by iteratively guiding an initial state toward the solution through reconstruction guidance. Importantly, we introduce an advanced inverse problem solver specifically designed for separation, which mitigates gradient conflicts caused by interference between the diffusion prior and reconstruction guidance during inverse denoising. This design ensures high-quality and balanced separation performance across individual sources. Additionally, we find that initializing the denoising process with an augmented mixture instead of pure Gaussian noise provides an informative starting point that significantly improves the final performance. To further enhance audio prior modeling, we design a novel time–frequency attention-based network architecture that demonstrates strong audio modeling capability. Collectively, these improvements lead to significant performance gains, as validated across speech–sound event, sound event, and speech separation tasks. 
\end{abstract}

\begin{links}
    \link{Project Page}{https://runwushi.github.io/unasdiff/}
\end{links}

\section{Introduction}
\begin{figure}[t]
\centering
\includegraphics[width=0.47\textwidth]{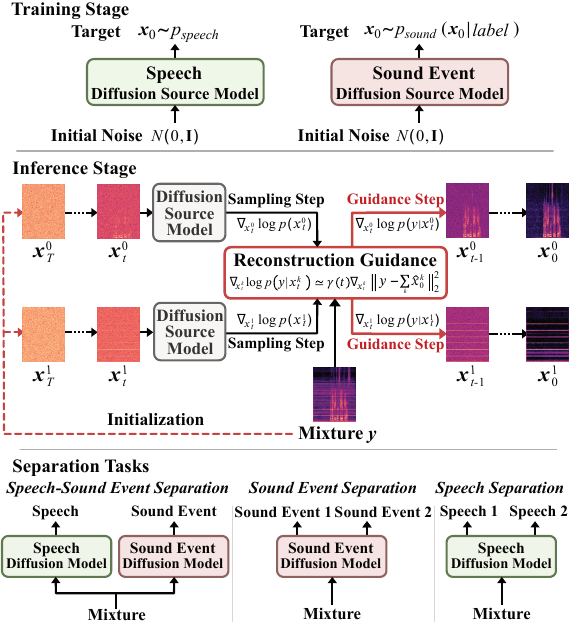}
\caption{Diffusion source model-based audio separation.}
\label{fig1}
\end{figure}
Single-channel audio separation aims to recover individual sources from a single-channel mixture and has long been an active research topic ~\cite{makino2018audio}. This problem is ill-posed with non-unique solutions and is typically addressed in a supervised manner, which requires large amounts of clean sources and synthetic mixtures for training. This limitation is particularly evident as recent research increasingly focuses on universal, multi-category audio separation tasks, encompassing speech, environmental sound events (e.g., keyboard keystrokes, animal vocalizations), and varied background sounds \cite{ma2024clapsep, saijo2025task}. Such tasks demand extensive training mixtures to represent diverse source combinations, which leads to a higher demand for data and training resources. To address this issue, we consider a more challenging setting without paired data, where the model learns from unpaired individual source data and separates audio signals through training-free guidance.

Traditional approaches for unpaired audio separation often model individual sound sources within the source model-based paradigm. Specifically, they characterized the intrinsic properties of each source signal to enable separation during inference. Such methods include Non-negative Matrix Factorization (NMF) \cite{le2015deep}, Hidden Markov Model (HMM) \cite{wang2014informed}, and Bayesian models \cite{benaroya2005audio}. Crucially, the priors inherent in such source modeling can be powerfully realized through modern deep generative models, including variational autoencoders (VAE) \cite{karamatli2019audio}, Glow-based generative models \cite{zhu2022music}, and generative adversarial networks (GAN) \cite{narayanaswamy2020unsupervised}. While promising, these techniques encounter performance ceilings inherent in the modeling capabilities and generative capacity of their underlying source models, significantly impacting separation performance. More recently, diffusion models \cite{cao2024survey} have shown unprecedented generative power in a wide range of audio tasks such as text-to-speech \cite{jung2025voicedit}. Moreover, their iterative denoising inference supplies strong data priors, making diffusion models well-suited for inverse problems, which aim to recover an unknown signal $\bm{x}$ from noisy measurement $\bm{y}$. 

To this end, we introduce our unsupervised diffusion-based audio separation framework, as illustrated in Figure~\ref{fig1}. We treat separation as a specific instance of the diffusion inverse problem, which involves two main considerations. On the one hand, when directly applying classical gradient descent based inverse-problem solvers such as Diffusion Posterior Sampling (DPS) \cite{chung2023diffusion} and Diffusion with Spherical Gaussian constraint (DSG) \cite{yang2024guidance} to audio separation, we observe the diffusion prior $\nabla_{\bm{x}_t}\log p(\bm{x}_t)$ and the reconstruction guidance $\nabla_{\bm{x}_t}\mathcal{L}_{\text{recon}}(\bm{y},\hat{\bm{y}})$ suffer from severe gradient conflicts. These conflicts degrade separation quality and hinder accurate source recovery, often resulting in noisy or incomplete outputs. On the other hand, widely adopted representations like mel-spectrograms for audio diffusion models are nonlinear and therefore lack the additive property of time-domain waveforms. Existing waveform-based diffusion models \cite{kong2020diffwave}, even those incorporating self-attention mechanisms \cite{ku2025generative}, comparatively lack dedicated audio-specific designs for directly modeling fine-grained intra-frame and intra-subband features. This fundamentally limits their capacity to capture diverse frequency bands and long-range temporal dependencies in audio. Hence, a more powerful model that operates directly in the waveform space is urgently needed.

To address these issues, firstly, we propose a hybrid gradient guidance strength schedule during the reverse separation process, which mitigates gradient conflicts and substantially improves the separation quality. Secondly, we design a triple-path self-attention-based U-Net diffusion backbone as source models to perform guided sampling for the separation task, where the diffusion modeling process and separated audio are directly operated in the waveform space.  In addition, we find that initializing the diffusion process with a noise-augmented mixture waveform, rather than pure Gaussian noise, markedly boosts separation fidelity. To sum up, the main contributions of this work are as follows:

\begin{itemize}
\item We are the first to address unpaired, general audio separation with diffusion models, framing the task as an inverse problem solved via flexible unconditional or conditional source models, supporting various separation tasks.
\item We analyze gradient conflicts and suboptimal prior initialization in the inverse separation process, proposing an effective guidance strength schedule and noise-augmented mixture initialization for separation.
\item We present a novel TF domain diffusion model design, employing a triple-path self-attention mechanism to effectively learn priors for diverse audio types like speech and sound events.
\item Experiments across speech–sound event, sound event, and speech separation tasks on VCTK and FSD-Kaggle2018 datasets demonstrate the superior separation quality of our proposed method, even achieving performance comparable to fully supervised models.
\end{itemize}

\section{Related Works}
\subsection{Single Channel Audio Separation}
Audio separation can be broadly categorized into supervised and unsupervised methods. Supervised techniques typically learn to separate sources from paired mixture-source data, often employing Permutation Invariant Training (PIT) to handle output ambiguity \cite{yu2017permutation}. These methods can operate in either the time-domain \cite{luo2019conv, li2022efficient} or frequency-domain \cite{wang2023tf} to separate audio signals.

Many unsupervised approaches perform source model-based separation using deep generative models. These approaches typically model the prior distribution of individual source signals and then infer the posterior distribution of sources given an observed mixture. In \cite{narayanaswamy2020unsupervised}, source-specific GAN generators served as audio source priors. Separation was achieved by iteratively optimizing latent codes via gradient descent on a waveform reconstruction loss. \cite{zhu2022music} proposed music separation using source-specific Glow models, similarly employing iterative gradient descent to optimize latent codes and recover sources. \cite{jayaram2020source} introduced a Bayesian approach for image separation using score-based generative models and Glow models as source priors. Source estimates were obtained by sampling from the posterior via noise-annealed Langevin dynamics. \cite{postolache2023latent} used pretrained VQ-VAEs to define source priors for image and audio separation, estimating posterior probability via direct statistical computation of dataset co-occurrences. \cite{mariani2023multi} applied diffusion models to music separation by modeling the joint prior distribution of sources and performing posterior sampling conditioned on the mixture.

\subsection{Diffusion Inverse Problems and Audio Applications}
Inverse problems can be modeled as $ \bm{y} = A(\bm{x}) + \bm{n}$, where $\bm{y}$ is the measurement, $A$ is the degradation model, $\bm{x}$ is the unknown underlying signal, and $\bm{n}$ is measurement noise. 
These problems are fundamental across many fields, enabling tasks such as image restoration \cite{fei2023generative} and audio restoration tasks \cite{iashchenko2023undiff}. Many diffusion-based inverse problem solvers, such as DPS \cite{chung2023diffusion}, work by iteratively refining samples through gradient descent. DPS is a well-known framework that tackles diverse inverse problems. Building upon this, DSG \cite{yang2024guidance} refines DPS. DSG optimizes the guidance by choosing a theoretically optimal step size \cite{daras2024survey}, combining it with projected gradient descent to achieve superior sample quality, particularly for high-dimensional data. For audio inverse problems, \cite{moliner2023solving} adopts a DPS-style method and applies it to bandwidth extension, audio inpainting, and declipping tasks. \cite{xu2025unsupervised} solves multi-channel speech separation using Independent Vector Analysis (IVA) for initialization and then applies a DPS-style method to achieve the separation.

\section{Methodology}
\subsection{Score-based Diffusion Models}
Score-based diffusion models are used to learn audio source priors, which involves a forward noising process where a clean audio sample $\bm{x}_t$ gradually becomes pure noise over time $t \in [0, T]$ \cite{song2020score}:
\begin{equation}
    \mathrm{d}\bm{x}= -\frac{\beta(t)}{2}\,\bm{x}\,\mathrm{d}t + \sqrt{\beta(t)}\,\mathrm{d}\bm{w},
\end{equation}
where $\beta(t)$ is the noise schedule and $\bm{w}$ is the Wiener process.
The corresponding reverse process is:
\begin{equation}
    \mathrm{d}\bm{x} = \left[-\frac{\beta(t)}{2}\bm{x} - \beta(t){\nabla_{\bm{x}_t}}\log
    p_{t}(\bm{x}_t)\right]\mathrm{d}t + \sqrt{\beta(t)}\mathrm{d}\bm{\bar{w}},
\end{equation}
where $\bar{\bm{w}}$ is the Wiener process in backward. The score function ${\nabla_{\bm{x}_t}}\log p_{t}(\bm{x}_t)$ is parametrically represented by a neural network $s_{\theta}$, trained via score-based denoising objectives. This provides differentiable estimation for single-source distributions, forming the prior models for separation tasks.

\subsection{Audio Separation as a Diffusion Inverse Problem}
The audio separation task can be formulated as an inverse problem: a linear mixing operator \(A\) combines the \(K\) sources and additive noise to produce the observed mixture,
\begin{equation}
\bm y \;=\; A(\bm{x}^1,\dots,\bm{x}^K)+\bm{n}\;=\;\sum_{k=1}^{K}\bm{x}^k + \bm n ,
\end{equation}
and the goal is to recover the individual sources \(\{\bm{x}^k\}_{k=1}^{K}\) from the single observation $\bm{y}$.

Throughout the diffusion process, we write \(\bm{x}_{t}^{k}\) for the state of source $k$ at timestep $t$.
Probabilistically, this corresponds to sampling from the posterior
\(p(\bm{x}^{1:K}|\bm{y})\).
Within the score-based framework, we realize this by guiding the reverse diffusion process~(2) with the conditional score
\(\nabla_{ \bm{x}_t^k}\log p(\bm{x}_t^k|\bm{y})\),
which is decomposed by Bayes’ rule as:
\begin{equation}
\nabla_{\bm{x}_t^k} \log p(\bm{x}_t^k | \bm{y}) = 
\nabla_{\bm{x}_t^k} \log p(\bm{x}_t^k) + \nabla_{\bm{x}_t^k} \log p(\bm{y}|\bm{x}_t^k),
\end{equation}
where $\nabla_{\bm{x}_t^k} \log p(\bm{x}_t^k)$ is obtained by the trained diffusion source model, and $\nabla_{\bm{x}_t^k} \log p(\bm{y}|\bm{x}_t^k)$ is the gradient of the log-likelihood. Following DPS, the likelihood term $\nabla_{\bm{x}_t^k} \log p(\bm{y}|\bm{x}_t^k)$ is approximated  by replacing it with  $\nabla_{\bm{x}_t^k} \log p(\bm{y}|\hat{\bm{x}}_0^k)$, where $\hat{\bm{x}}_0^k=\mathbb{E}[\bm{x}_0^k|\bm{x}_t^k]$ is the estimated clean sample computed via Tweedie’s formula \cite{efron2011tweedie}:  
\begin{equation}
\hat{\bm{x}}_0^k \simeq (\bm{x}_t^k+(1-\bar{\alpha}(t))\bm{s}_{\theta}^k(\bm{x}_t^k, t))/{\sqrt{\bar{\alpha}(t)}},
\end{equation}
where $\bar{\alpha}(t) = \prod_{j=1}^t(1-\beta{(j)})$ is the cumulative product of the discrete variance schedule, $s_{\theta}^k$ is the $k$ th source model. With the estimated clean sources $\hat{\bm{x}}_{0}^k$, and under a Gaussian assumption for $p(\bm{y}|\sum_{k=1}^{K}\bm{x}_{0}^k)$, the likelihood gradient for $k$ th source can be approximated through backpropagation:
\begin{equation}
\nabla_{\bm{x}_{t}^k} \log p(\bm{y}|\bm{x}_{t}^k)
\;\simeq\;
\gamma(t)\,
\nabla_{\bm{x}_{t}^k} \mathcal{L}_\text{recons}\bigl(\bm{y}, \hat{\bm{y}}\bigr),
\end{equation}
where $\hat{\bm{y}} = \sum_{k=1}^{K} \hat{\bm{x}}_{0}^k$ is the reconstructed mixture, and $\gamma{(t)}$ is a step-dependent coefficient that controls the guidance strength. The design of this schedule is crucial to the final separation performance, as shown in our experiments. 
The loss function $\mathcal{L}_\text{recons}\bigl(\bm{y}, \hat{\bm{y}}\bigr)$, utilized for computing guidance gradients, is a weighted sum of reconstruction errors in both the time and frequency domains, which is defined as:
\[
\mathcal{L}_{\text{recons}}
   =\lambda_{\text{time}}\mathcal{L}_{\text{time}}
   +\lambda_{\text{group}}\mathcal{L}_{\text{group}}
   +\lambda_{\text{stft}}\mathcal{L}_{\text{stft}},
\]
where
\(\mathcal{L}_{\text{time}}=\lVert\bm y-\hat{\bm y}\rVert_{2}^2\) measures the reconstruction error in time domain, $\mathcal{L}_{\text{group}}=(\sum_{n}\lVert\bm y_{(n)}-\hat{\bm y}_{(n)}\rVert_{2}^2)/{N_g}$ averages the same error over $N_g$ non-overlapping segments indexed by $n$, and $\mathcal{L}_{\text{stft}}=\lVert\,|\,\text{STFT}(\bm y)|-|\,\text{STFT}(\hat{\bm y})|\rVert_{2}^2$ compares magnitude spectrogram in the Short-time Fourier transform (STFT) domain. The algorithm is presented in Algorithm \ref{alg1}.

\begin{algorithm}
    [tb]
    \caption{Audio Separation with Diffusion Priors}
    \label{alg1} 
    \textbf{Require}: number of sound sources $K$, diffusion time step $T$, noise levels $\{\sigma_{j}\}_{j=1}^{T}$, gradient guidance schedule $\{\gamma_{j}\}_{j=1}^{T}$, initialization step $t^*$, diffusion source models $\{s_{\theta}^k\}_{k=1}^K$ \\
    \textbf{Input:} mixture $\bm{y}$, sound event label $\bm{c}$ (optional)

    \begin{algorithmic}[1] 
    \STATE \colorbox{gray!15}{\text{Initialization from mixture:}}
        \STATE $
            \{\bm{x}_{T}^\text{aug}\}_{k=1}^K = \sqrt{\bar{\alpha}_{t^{*}}}\bm{y} + \sqrt{1-\bar{\alpha}_{t^{*}}}\bm{\epsilon}, \quad \bm{\epsilon} \sim \mathcal{N}(0, \mathbf{I})
            $
        
      \FOR{$t = T-1$ \TO $0$ }
      \FOR{$k = 1$ \TO $K$}
        \STATE $\bm{\epsilon}_t \sim \mathcal{N}(0, \mathbf{I})$
        \STATE $\hat{\bm{x}}_{0}^k \gets (\bm{x_t}+(1-\bar{\alpha}_t)\bm{s}_{\theta}^k(\bm{x}_{t}^k, t,[\bm{c}]))/{\sqrt{\bar{\alpha}_t}}$
        \STATE \colorbox{gray!15}{\text{Prior sampling:}}
        
          \STATE $\bm{x}_{t-1}^{k'}\leftarrow \frac{\sqrt{\alpha_t}(1-\bar\alpha_{t-1})}{1-\bar\alpha_t}\,\bm{x}_{t}^k
            + \frac{\sqrt{\bar\alpha_{t-1}}\beta_t}{1-\bar\alpha_t}\,\hat{\bm{x}}_{0}^k
            + \sigma_t\,\bm{\epsilon}_t$
      \ENDFOR
      
      \STATE $\hat{\bm{y}} = \sum_{k=1}^{K} \hat{\bm{x}}_{0}^k$
      \STATE     $\mathcal{L}_{\text{recons}}(\bm{y}, \hat{\bm{y}}) = \lambda_{\text{time}}\mathcal{L}_{\text{time}} + \lambda_{\text{group}}\mathcal{L}_{\text{group}} + \lambda_{\text{stft}}\mathcal{L}_{\text{stft}}$
      
  \FOR{$k = 1$ \TO $K$}
    \STATE \colorbox{gray!15}{\text{Back propagation:}}
    
    \STATE $\nabla_{\bm{x}_{t}^k} \log p(\bm{y}|\bm{x}_{t}^k) \;\simeq\; \gamma(t) \nabla_{\bm{x}_{t}^k} \mathcal{L}_\text{recons}\bigl(\bm{y}, \hat{\bm{y}}\bigr)$
      \STATE $\bm{x}_{t-1}^k \leftarrow \bm{x}_{t-1}^{k'}-\nabla_{\bm{x}_{t}^k} \log p(\bm{y}|\bm{x}_{t}^k)$
  \ENDFOR
      \ENDFOR
       \STATE \textbf{return} $\{\bm{x}_{0}^k\}_{k=1}^K$ 
    \end{algorithmic}
\end{algorithm}

\begin{figure*}[t]
\centering
\includegraphics[width=1.0\textwidth]{figs/02.pdf} 
\caption{(a) Constant guidance of DPS. (b) Noise-proportional guidance of DSG. (c) Our proposed hybrid guidance schedule. The first row shows the gradient conflict metrics of all sources. The second row shows the SI-SDR of all sources. The third row visualizes the mixtures, intermediate states, and the ground truth of the sources. }
\label{fig2}
\end{figure*}

\subsection{Design of Guidance Strength Schedule $\gamma(t)$}
The performance of solving the separation problem is critically influenced by the design of the guidance schedule $\gamma(t)$, which weights the strength of gradient updates. While various guidance strategies exist as presented in Table~\ref{tab:1}, some are designed for general inverse problems and may not be optimal for the specific constraints of audio separation.

To gain deeper insights, we analyze the generative behavior of sources from the perspective of gradient updates. The overall inverse sampling process can be viewed as a multi-objective optimization problem, as discussed in \cite{yu2020gradient}. This involves two primary objectives:
\begin{enumerate}
    \item \textbf{Prior Objective:} $\nabla_{\bm{x}_t} \log p(\bm{x}_t)$, aiming to maintain the learned structure and ensuring the generated sample remains on the manifold of natural data, leading to a prior gradient $g_\text{prior}(t)$.
    \item \textbf{Likelihood Objective:} $\nabla_{\bm{x}_t} \log p(\bm{y}|\bm{x}_t)$, ensuring the sample adheres to the physical constraints $\|\bm{y}-\sum_{k=1}^K\hat{\bm{x}}_0^k\|_2$ imposed by the measurement $\bm{y}$, resulting in a conditional guidance gradient $g_\text{cond}(t)$.
\end{enumerate}
Consequently, the total gradient can be expressed as:
\begin{equation}
   \nabla_{\bm{x}_t} \log p(\bm{x}_t | \bm{y}) \propto g_\text{prior}(t) + \gamma(t)g_\text{cond}(t).
\end{equation}
To minimize the reconstruction loss $\mathcal{L}_{\text{recons}}$, we require the update direction to have a positive projection onto the conditional gradient $g_{\text{cond}}(t)$:
\begin{equation}
   (\nabla_{\bm{x}_t} \log p(\bm{x}_t | \bm{y}))^\top g_{\text{cond}}(t) > 0.
\end{equation}
Substituting the update formula and simplifying yields the following condition for making progress:
\begin{equation}
   \gamma(t) > -\frac{{g_{\text{prior}}(t)}^\top {g_{\text{cond}}(t)}}{\| {g_{\text{cond}}(t)} \|^2}.
\end{equation}
We define the guidance bound metric $g_{\text{bound}}(t) := -\frac{{g_{\text{prior}}(t)}^\top {g_{\text{cond}}(t)}}{\| {g_{\text{cond}}(t)} \|^2}$. A positive $g_{\text{bound}}(t)$ indicates a gradient conflict between the prior gradient $g_{\text{prior}}(t)$ and the conditional gradient ${g_{\text{cond}}(t)}$, occurring when their dot product is negative, indicating opposing directions. 
For effective progress, the guidance strength $\gamma(t)$ should exceed this bound to overcome the conflict.

As illustrated in Figure~\ref{fig2}, we evaluated the separation process of a 4-second audio mixture (fiddle and chime from the FSD test dataset) using different guidance strategies: DPS, DSG, and our proposed hybrid schedule. The top row of Figure~\ref{fig2} illustrates our guidance conflict metric $g_{\text{bound}}(t)$, while the second row shows the Scale-Invariant Signal Distortion Ratio (SI-SDR) for each source across diffusion steps. We observed that constant guidance DPS often faces significant gradient conflicts in early, high-noise stages (Figure~\ref{fig2}(a)), requiring stronger initial guidance. Conversely, noise-proportional guidance DSG effectively mitigates these early conflicts, but its diminishing step size leads to rising gradient conflict and imbalanced separation in the final, low-noise regime (Figure~\ref{fig2}(b)).

These complementary trade-offs, consistently observed across samples, directly inspire our novel hybrid guidance strength schedule. Our approach combines the early-stage adaptability of the noise-proportional schedule with the late-stage stability of the constant one. This is achieved using the $\operatorname{SmoothMax}$ function:
\begin{equation}
\operatorname{SmoothMax}_c(a, b) = \frac{1}{c} \log\left( \exp(c a) + \exp(c b) \right),
\end{equation}
which defines our proposed gradient strength schedule as:
\begin{equation}
\gamma(t) = \frac{\operatorname{SmoothMax}_c\left( \sigma(t),\; s_{\text{floor}} \right)\sqrt{N}}{\|{\nabla_{\bm{x}_{t}}\mathcal{L}_\text{recons}}\| _2},
\end{equation}
where $ \sigma(t) $ is the noise-dependent standard deviation, defined as $\sqrt{\beta(t){(1-\bar{\alpha}(t-1))}/{1-\bar{\alpha}(t)}} $. $N$ is the signal length. $ s_{\text{floor}}$ is a constant that enforces a non-vanishing lower bound, and $c$ controls the sharpness of the transition. 
Different schedules without the normalization term and $\sqrt{N}$ are presented in Figure~\ref{fig3}. 
By integrating these complementary strengths, our hybrid approach can enhance separation quality and source balance, as shown in Figure~\ref{fig2}(c).

\begin{figure}[h] 
    \centering
    \includegraphics[width=0.9\linewidth]{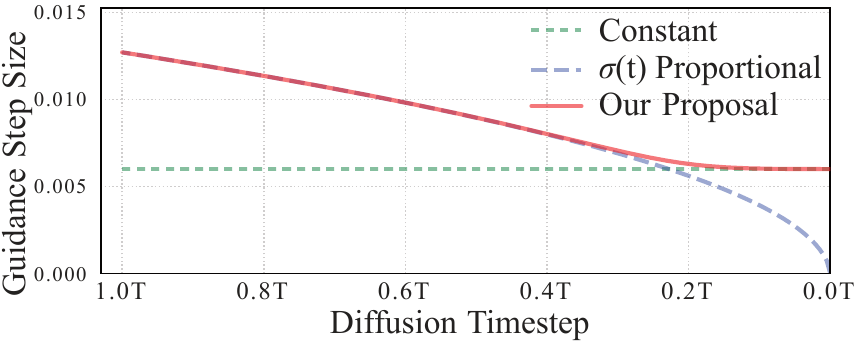}
    \caption{Different guidance strength schedules.}
    \label{fig3}
\end{figure}

\begin{table}[t]
\centering
\small 
\begin{tabular}{
    m{4.5cm} 
    >{\centering\arraybackslash}m{3.2cm}
}
\toprule
\textbf{Guidance Strength Strategy $\gamma(t)$ } & \textbf{Formulation} \\
\midrule
\textbf{DPS} \cite{chung2023diffusion}: Constant. 
& 
$ \gamma(t) = \text{const} $ 
\\
\addlinespace
\textbf{DSG} \cite{yang2024guidance}: Proportional to the noise level $\sigma(t)$ .
& 
$
\gamma(t) = \frac{\sigma(t)\sqrt{N}}{\|{\nabla_{\bm{x}_{t}}\mathcal{L}_\text{recons}}\| _2}
$
\\
\addlinespace
\textbf{Our proposed}: First proportional to the noise level $\sigma_t$, then smoothly transformed to a constant guidance.
&
$
\frac{\operatorname{SmoothMax}_c\bigl(\sigma(t),\; s_{\text{floor}}\bigr)\,\sqrt{N}}
           {\bigl\lVert \nabla_{\bm{x}_{t}} \mathcal{L}_{\text{recons}} \bigr\rVert_{2}}
$
\\
\addlinespace
\bottomrule
\end{tabular}
\caption{Comparison of guidance strength schedule.}
\label{tab:1}
\end{table}

\subsection{Improve Separation Fidelity with Initialization}
In diffusion-based inverse problems, most methods start sampling from pure Gaussian noise \cite{chung2023diffusion, yang2024guidance}. However, this is suboptimal for audio separation, where sources exhibit strong structure and are constrained by their mixture composition. Starting from noise forces the model to explore a vast, unstructured space, essentially recovering everything from scratch, which often limits the separation quality.

To address this, we adopt a single-step initialization \cite{chung2022come}: instead of pure noise, each source is seeded by the same noised version of the mixture :
\begin{equation}
\bm{x}_T^\text{aug} = \sqrt{\bar{\alpha}_{t^{*}}}\bm{y} + \sqrt{1-\bar{\alpha}_{t^{*}}}\bm{\epsilon}, \quad \bm{\epsilon} \sim \mathcal{N}(0, \mathbf{I})
\end{equation}
where $t^{*}$ is a chosen time step that controls the initial noise variance. This single latent state $\bm{x}_{T}^\text{aug} $ thus serves as a common starting point for all source channels, grounding the subsequent separation process in a shared, coherent context derived directly from the measurement $\bm{y}$. The utilized single-step, analytical initialization is different from the DDPM Inversion \cite{liu2024accelerating}, which requires an iterative sampling process and whose performance depends on the learned model.

\subsection{Model Architecture Design}
\begin{figure}
    \centering
    \includegraphics[width=1.0\linewidth]{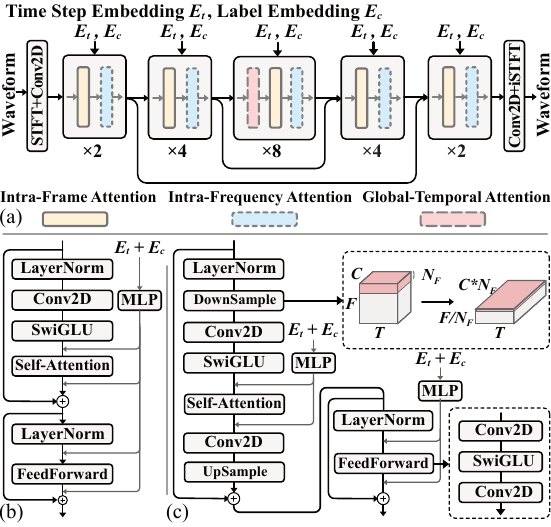}
      \caption{(a) Overall architecture of the proposed model. (b) Structure of the Intra‑Frame Attention and Intra‑Frequency Attention blocks. (c) Global‑Temporal Attention block.}
    \label{fig4}
\end{figure}

 Current diffusion audio backbones typically lack fine-grained acoustic modeling with dedicated attention mechanisms, particularly in capturing inter-frame and sub-band interactions, for which we design a triple-path fully attention-based U-Net diffusion backbone that directly predicts complex-valued noise from STFT spectrograms, as shown in Figure~\ref{fig4}(a). Its core is three kinds of attention blocks that capture distinct spectro-temporal correlations. Each block uses learnable Swish Gated Linear Unit (SwiGLU)~\cite{shazeer2020glu} for both self-attention preprocessing and feed-forward layers. The entire network is conditioned using a timestep embedding $E_t$ obtained via sinusoidal encoding followed by a multilayer perceptron (MLP)~\cite{tian2024u}. For sound event modeling, an additional class vector $E_c$ from a learnable embedding dictionary corresponding to sound class labels is also integrated. Both $E_t$ and $E_c$ are incorporated via the AdaLN-Zero mechanism~\cite{peebles2023scalable}.

\noindent
\textbf{Intra‐Frame and Intra‐Frequency Attention} 
As shown in Figure~\ref{fig4}(b), this attention strategy models inter-frequency correlations within each frame and temporal dynamics of individual frequency bands separately. Intra-Frame Attention scans across frequency bins within each time frame, while Intra-Frequency Attention focuses on temporal evolution within each frequency bin, collectively capturing fine-grained spectral and temporal features. These two blocks share the same structure while differing in scan direction. This approach has been effectively adopted in various speech processing models \cite{luo2020dual, shi2025distance, xu2024tiger}.

\noindent
\textbf{Global‐Temporal Attention} The Global-Temporal Attention block, illustrated in Figure~\ref{fig4}(c), is designed to capture temporal relationships across all frequency bins. To reduce computational cost, the input feature map is downsampled along the frequency dimension and processed by a SwiGLU module for channel projection. After attention calculation, a convolution layer projects the feature map back. This block is exclusively used in the latent stage of the U-Net.

\section{Experiments and Results}
\subsection{Datesets and Preprocessing}
We independently train diffusion source models for speech and sound event datasets. For speech modeling, we adopt the VCTK dataset, which contains approximately 110 speakers, using 100 speakers for training and 10 for testing. For sound event modeling, we utilize the FSD-Kaggle2018 dataset~\cite{fonseca2018general}, which comprises around 11,000 clips across 41 event categories. We use the provided training set with 9,470 clips for training and the remaining clips for testing. All audio samples are resampled to 16 kHz and are cropped or padded to a fixed length of 4 seconds.

\subsection{Configuration and Training}
For the U-Net backbone, we adopt an STFT with a window size of 510 and a hop length of 255. The channel size is set to 72. For Global-Temporal Attention, the number of subbands is set to 4, and the suppressed channel is set to 16. Each attention block employs 4 heads, and the embedding dimension is 128. As illustrated in Figure~\ref{fig4}, the U-Net adopts a five-stage block layout with repetitions of 2, 4, 8, 4, and 2, and the total number of parameters is 37M. Each model is trained for 400k steps with a batch size of 12 using the AdamW optimizer with a learning rate of 1e-4~\cite{loshchilov2017decoupled}. We train two diffusion source models on speech and sound event datasets. 
Denoising training is performed with $T = 200$ steps using a linear noise schedule $\beta(t) \in [10^{-4}, 2 \times 10^{-2}]$. $s_\text{floor}$ is set to 0.002 and $c$ is set to $10^3$. $\lambda_{\text{time}}$, $\lambda_{\text{group}}$, and $\lambda_{\text{stft}}$ are set to 1, 0.05, and 0.1, respectively. The initialization step $t^*$ is set to 150.

\subsection{Experimental Implementation}
\subsubsection{Mixture Generation and Tasks}
For each of the three distinct tasks investigated: Speech-Sound Event Separation, Sound Event Separation, and Speech Separation, 400 unique mixtures were synthesized. These mixtures were created by overlapping single sources with random Root Mean Square levels (-25 to -20 dB) and random temporal offsets.

For Speech-Sound Event Separation, we tested separating speech from sound events in mixtures containing one speech source with either one or two sound event sources, utilizing an unconditional speech model and a label-guided conditional sound event model. Sound Event Separation focused on separating two or three different sound types within mixtures, employing a single conditional sound event model.

For speech separation, a homogeneous source separation problem involves two speech sources. This scenario is particularly challenging for source model-based methods, since it inherently presents permutation ambiguity, as local speech segments may separate, but consistent speaker assignment across time often fails because models trained on clean speech lack speaker clues and primarily learn local acoustic structures \cite{iashchenko2023undiff}. Our results suggest that a stronger model design can enhance such homogeneous separation without relying on explicit speaker conditioning.

\subsubsection{Comparison Methodologies and Metrics}
Comparisons are made against two supervised separation models: the classical time-domain Conv-TasNet \cite{luo2019conv}, and the state-of-the-art frequency-domain audio separation model TF-Locoformer \cite{saijo2024tf, saijo2025task}. We independently train the corresponding expert model with the SI-SDR loss function for 200 epochs for each task. Across all comparisons, we utilize identical mixture test datasets.

For unsupervised methods, we evaluate our proposed hybrid guidance schedule $\gamma(t)$ against DPS's constant and DSG's noise-proportional schedule, all within the gradient-descent framework and using the same initialization. Additionally, we compare with the Analytical Sampling method \cite{iashchenko2023undiff}. This approach is used for additive source separation, where the observed mixture $\bm{y}$ is an additive combination of sources $\bm{y}=\bm{x}_1+\bm{x}_2$. It leverages the key insight that the conditional likelihood $p_t(\bm{y}|\bm{x}_{t}^1,\bm{x}_{t}^2)$ simplifies to $p_t(\bm{y}|\bm{x}_{t}^1+\bm{x}_{t}^2)$. This enables the analytical computation of the log-likelihood gradient $\nabla_{\bm{x}_{t}^k}\log p_t(\bm{y}|\bm{x}_{t}^1,\bm{x}_{t}^2)$.

For Speech Separation specifically, our full attention-based model is compared with large Diffwave \cite{iashchenko2023undiff, kong2020diffwave}. This enhanced Diffwave model incorporates the Squeeze-Excitation mechanism, with 33M parameters and 48 layers. We use its publicly released checkpoint trained on 2-second VCTK segments to compare its speech modeling performance with ours.

Performance is evaluated using standard objective metrics: SI-SDR for all separated sources, and PESQ for separated speech in Speech-Sound and Speech separation tasks. For Speech Separation specifically, we also report the separation failure rate. Failure is defined as instances where the average SI-SDR of separated sources falls below 0 dB. 

\begin{table*}[t]
  \centering
  \setlength{\tabcolsep}{2pt} 
  
  \begin{tabular*}{\linewidth}{ %
      >{\raggedright\arraybackslash}p{4cm} 
      | 
      @{\hspace{1.0em}} 
      >{\centering\arraybackslash}p{1.5cm} %
      @{\hspace{0.4em}} 
      >{\centering\arraybackslash}p{1.2cm} 
      @{\hspace{0.4em}}|@{\hspace{0.4em}} 
      >{\centering\arraybackslash}p{1.5cm} 
      @{\hspace{0.4em}}
      >{\centering\arraybackslash}p{1.2cm} 
      @{\hspace{0.4em}}|@{\hspace{0.4em}} 
      >{\centering\arraybackslash}p{1.5cm} 
      @{\hspace{0.4em}}|@{\hspace{0.4em}} 
      >{\centering\arraybackslash}p{1.5cm} 
      @{\hspace{0.4em}}|@{\hspace{0.4em}} 
      >{\centering\arraybackslash}p{1.3cm} 
      @{\hspace{0.4em}} 
      >{\centering\arraybackslash}p{1.0cm} 
      @{\hspace{0.4em}} 
      >{\centering\arraybackslash}p{1.2cm} 
      }
    \multicolumn{1}{c}{} & 
      \multicolumn{2}{c}{\textbf{1 Speech + 1 Sound}} & 
      \multicolumn{2}{c}{\textbf{1 Speech + 2 Sound}} & 
      \multicolumn{1}{c}{\textbf{2 Sound}} &
      \multicolumn{1}{c}{\textbf{3 Sound}} &
      \multicolumn{3}{c}{\textbf{2 Speech}} \\
    \toprule

    \multicolumn{1}{>{\centering\arraybackslash}p{3.5cm}!{\vrule}}{\textbf{Method}} & 
      \textbf{SI-SDR} ($\uparrow$) & \textbf{PESQ} ($\uparrow$) &
      \textbf{SI-SDR} ($\uparrow$) & \textbf{PESQ} ($\uparrow$) &
      \textbf{SI-SDR} ($\uparrow$) & \textbf{SI-SDR} ($\uparrow$)
      & \textbf{SI-SDR}  ($\uparrow$)
      & \textbf{PESQ} ($\uparrow$)
      & \textbf{Failure Rate} ($\downarrow$)
      \\ 
    \midrule
    \multicolumn{1}{l|}{{Unprocessed}} & 
      0.00 & 1.64 & -3.15 & 1.22 & -0.01 & -3.13 & 0.01 & 1.29 & -- \\ 
    \midrule
    
    \multicolumn{5}{l}{\textit{Unsupervised: Gradient+Initialization}}
    \\
    \multicolumn{1}{l|}{\hspace{1em}DPS's constant $\gamma(t)$} & 8.88 & 1.73 & 3.02 & 1.56 & 4.63 & 0.12 & 1.87 & 1.43 & 46.5\%  \\ 
    \multicolumn{1}{l|}{\hspace{1em}DSG's $\sigma(t)$-prop. $\gamma(t)$} & 12.97 & 1.94 & 7.96 & 1.72 & 9.48 & 4.75 & 5.68 & 1.57 & 31.3\% \\ 
    \multicolumn{1}{l|}{\hspace{1em}\textbf{Proposed $\gamma(t)$}} & \underline{\textbf{14.31}} & \underline{\textbf{2.12}} & 
    \underline{\textbf{8.58}} & \underline{\textbf{1.80}} & \underline{\textbf{10.44}} & \underline{\textbf{5.15}} & \underline{\textbf{6.11}} & \underline{\textbf{1.62}} & \underline{\textbf{29.8\%}}  \\ 
    \midrule

    \multicolumn{4}{l}{{\textit{Unsupervised: Analytical Sampling}}} 
    \\[1pt]
\multicolumn{1}{>{\raggedright}p{3.5cm}|}{%
    \hspace{1em}\footnotesize \cite{iashchenko2023undiff}  
} 
& 7.33 & 1.45 & -0.04 & 1.19 & 1.01 & -2.18 & 2.17 & 1.28 & 42.3\% \\  %
    \midrule
    
    \multicolumn{1}{l}{\textit{Supervised}} \\
\multicolumn{1}{>{\raggedright}p{3.75cm}|}{%
    \hspace{1.0em}Conv-TasNet  
} 
    & 14.22 & 1.97 & 7.05 & 1.43 & 11.73 & 5.02 & 9.63 & 1.80 & --  \\ \multicolumn{1}{>{\raggedright}p{3.75cm}|}{%
    \hspace{1.0em}TF-Locoformer  
} 
    & \textbf{18.22} & \textbf{2.37} & \textbf{12.21} & \textbf{1.95} & \textbf{14.32} & \textbf{9.27} & \textbf{14.52} & \textbf{2.42} & --  \\ 

    \bottomrule
  \end{tabular*}
\caption{Performance on three audio separation tasks: Speech-Sound Event Separation (1 Speech+1 Sound, 1 Speech+2 Sound), Sound Event Separation (2 Sound, 3 Sound), and Speech Separation (2 Speech). The failure rate is reported only for speech separation and is defined as the average SI-SDR of separated sources falling below 0 dB. \underline{Underline} indicates unsupervised best.}
  \label{tab:2}
\end{table*}

\begin{table}[t]
  \centering
  \begin{tabular}{ 
      >{\raggedright\arraybackslash}p{2.7cm} %
      >{\centering\arraybackslash}p{1.2cm} 
      >{\centering\arraybackslash}p{1.5cm} 
      >{\centering\arraybackslash}p{1.22cm} 
  }
    \toprule
    \multicolumn{1}{c}{\textbf{Method}} &
      \textbf{SI-SDR Succ}  ($\uparrow$)  & 
      \textbf{Failure Rate} ($\downarrow$) & \textbf{SI-SDR All}  ($\uparrow$) \\ 
    \midrule
    
    Unprocessed & -- & -- & 0.00/0.01  \\ 
    \midrule

    \multicolumn{4}{l}{\textit{Unsupervised: Gradient+Proposed $\gamma(t)$}}\\[1pt]
    \hspace{0.5em}Large Diffwave & 7.17/7.36 &  39\%/44\% & 3.37/3.01 \\  
    \hspace{0.5em}\text{Proposed Model}  & \textbf{9.34}/\textbf{9.68} & \textbf{28\%}/\textbf{30\%} & \textbf{6.04}/\textbf{6.11} \\

    \midrule
    \multicolumn{3}{l}{\textit{Unsupervised: Analytical sampling}}\\[1pt]
    \hspace{0.5em}Large Diffwave & 5.12/4.69 & 44\%/\textbf{41\%} & 2.01/1.99 \\ 
    \hspace{0.5em}\text{Proposed Model}  & \textbf{5.32/5.27} & \textbf{39\%}/42\% & \textbf{2.45/2.17} \\



    \bottomrule
  \end{tabular}
    \caption{Performance of Speech Separation using different backbones. All metrics report results for 2-second and 4-second audio segments, presented in a 2-second / 4-second format. SI-SDR Succ denotes successfully separated results.}
  \label{tab:3}
\end{table}

\subsection{Results}
Results for Speech-Sound Event Separation, Sound Event Separation, and Speech Separation are presented in Table~\ref{tab:2}. As anticipated, supervised methods achieve the highest separation quality, with TF-Locoformer demonstrating superior performance across all scenarios. Among the unsupervised approaches, our proposed hybrid guidance schedule consistently outperforms other gradient-based alternatives across all tasks, showcasing its robustness. Critically, for speech-containing tasks, our method yields significantly higher PESQ scores and lower failure rate, indicating enhanced speech naturalness. Notably, our proposed method yields highly competitive results when compared to supervised baselines, performing comparably to Conv-TasNet, and delivering significant improvements over the unprocessed baseline in all separation tasks. In general, gradient-based methods universally surpass Analytical sampling across tasks, particularly evident in the more challenging multi-source separation tasks. Overall, the results confirm that our schedule delivers superior performance in terms of both separation fidelity and speech naturalness.

\begin{table}[t]
  \centering
  \begin{tabular}{ 
      >{\raggedright\arraybackslash}p{2.5cm} 
      >{\centering\arraybackslash}p{2.3cm} 
      >{\centering\arraybackslash}p{2.2cm} 
  }
    \toprule
    \textbf{Parameters} & \textbf{SI-SDR} ($\uparrow$) & 
    \textbf{PESQ} ($\uparrow$)\\ 
    \midrule
    \multicolumn{3}{l}{\textit{Initialization step $t^*$}} \\ 
    \hspace{1em}200 (no $\bm{x}_T^\text{aug}$) & 10.21 (-4.10) & 1.90 (-0.22) \\ 
    \hspace{1em}175 & 13.73 (-0.58) & 2.10 (-0.02)\\ 
    \hspace{1em}125 & 14.31 (-0.00) & 2.11 (-0.01) \\
    \hspace{1em}100 & 14.07 (-0.24) & 2.10 (-0.02) \\
    \midrule
    \multicolumn{3}{l}{\textit{Loss Type of} {$\mathcal{L}_{\text{recons}}$}} \\ 
    \hspace{1em}$\mathcal{L}_{\text{time}}$ & 13.65 (-0.66) & 1.71 (-0.41) \\ 
    \hspace{1em}$\mathcal{L}_{\text{time}}+\mathcal{L}_{\text{group}}$ & 13.20 (-1.11) & 1.65 (-0.47)  \\ 
    \hspace{1em}$\mathcal{L}_{\text{time}}+\mathcal{L}_{\text{stft}}$ & 13.03 (-1.28) & 2.24 (+0.12) \\ 
    \midrule
    \multicolumn{3}{l}{\textit{Final guidance} $s_\text{floor}$} \\ 
    \hspace{1em}0.005 & 14.20 (-0.11) & 2.05 (-0.07) \\ 
    \hspace{1em}0.001 & 13.92 (-0.39) & 2.08 (-0.04) \\ 
    \bottomrule
  \end{tabular}
  \caption{Ablation study of Initialization and Guidance.}
  \label{tab:4}
\end{table}

\subsection{Ablation Studies}
\subsubsection{Influence of Diffusion Model Backbone}
This ablation study investigates the impact of the underlying diffusion model architecture on speech separation performance, as presented in Table~\ref{tab:3}. We compare our proposed attention-based model backbone against Large Diffwave using our gradient-based method and Analytical sampling.

Results are reported for both 2-second and 4-second speech segments. Our proposed model largely outperforms Large Diffwave in SI-SDR Success and Failure Rate across gradient-based and the Analytical sampling strategy. Our model was not specifically trained on 2-second segments, but its performance on these shorter durations remains superior to Large Diffwave. These findings indicate that our proposed diffusion model architecture is more effective at learning robust audio priors and facilitating superior separation, thereby significantly improving overall separation quality.

\subsubsection{Ablation of Initialization and Guidance Parameters}
Table \ref{tab:4} presents our ablation study on the key components for the Speech-Sound separation process. We first investigated the initialization time steps $t^*$. Initializing from pure noise without augmentation yielded the lowest performance, highlighting that an optimal intermediate $t^*$ is crucial, as both excessively high and low noise levels are suboptimal. Next, we examine the impact of reconstruction loss $\mathcal{L}_{\text{recons}}$, showing that combining time-domain and frequency-domain losses yields the best performance. Finally, we evaluated the influence of the guidance constant $s_\text{floor}$ in the final guidance stage. We observed that while even a minimal floor value $s_\text{floor}$ provides a clear gain, setting it to an intermediate value further enhances overall performance.

\section{Conclusion}
In this work, we address single-channel audio separation from an unsupervised perspective, framing it as a probabilistic inverse problem solvable with diffusion models. We observed and analyzed the phenomenon of gradient conflicts between prior and conditional updates. To effectively manage these conflicts, we proposed a hybrid guidance strength schedule. Furthermore, we introduce an effective noise-augmented mixture initialization strategy. These innovations significantly enhance separation quality and efficiency, built upon a novel TF domain model for robust audio priors learning. Our method demonstrates impressive effectiveness across three distinct audio separation tasks.

\cleardoublepage
\bibliography{aaai2026}

@article{ma2024clapsep,
  title={CLAPSep: Leveraging Contrastive Pre-trained Model for Multi-Modal Query-Conditioned Target Sound Extraction},
  author={Ma, Hao and Peng, Zhiyuan and Li, Xu and Shao, Mingjie and Wu, Xixin and Liu, Ju},
  journal={IEEE/ACM Transactions on Audio, Speech, and Language Processing},
  year={2024},
  publisher={IEEE}
}

@inproceedings{saijo2025task,
  title={Task-Aware Unified Source Separation},
  author={Saijo, Kohei and Ebbers, Janek and Germain, Fran{\c{c}}ois G and Wichern, Gordon and Le Roux, Jonathan},
  booktitle={ICASSP 2025-2025 IEEE International Conference on Acoustics, Speech and Signal Processing (ICASSP)},
  pages={1--5},
  year={2025},
  organization={IEEE}
}

@inproceedings{narayanaswamy2020unsupervised,
  title={Unsupervised Audio Source Separation Using Generative Priors},
  author={Narayanaswamy, Vivek and Thiagarajan, Jayaraman J and Anirudh, Rushil and Spanias, Andreas},
  booktitle={Proc. Interspeech 2020},
  pages={2657--2661},
  year={2020}
}

@article{zhu2022music,
  title={Music source separation with generative flow},
  author={Zhu, Ge and Darefsky, Jordan and Jiang, Fei and Selitskiy, Anton and Duan, Zhiyao},
  journal={IEEE Signal Processing Letters},
  volume={29},
  pages={2288--2292},
  year={2022},
  publisher={IEEE}
}

@inproceedings{postolache2023latent,
  title={Latent autoregressive source separation},
  author={Postolache, Emilian and Mariani, Giorgio and Mancusi, Michele and Santilli, Andrea and Cosmo, Luca and Rodol{\`a}, Emanuele},
  booktitle={Proceedings of the AAAI Conference on Artificial Intelligence},
  volume={37},
  number={8},
  pages={9444--9452},
  year={2023}
}

@inproceedings{jayaram2020source,
  title={Source separation with deep generative priors},
  author={Jayaram, Vivek and Thickstun, John},
  booktitle={International Conference on Machine Learning},
  pages={4724--4735},
  year={2020},
  organization={PMLR}
}

@article{fonseca2018general,
  title={General-purpose tagging of freesound audio with audioset labels: Task description, dataset, and baseline},
  author={Fonseca, Eduardo and Plakal, Manoj and Font, Frederic and Ellis, Daniel PW and Favory, Xavier and Pons, Jordi and Serra, Xavier},
  journal={arXiv preprint arXiv:1807.09902},
  year={2018}
}

@article{mariani2023multi,
  title={Multi-source diffusion models for simultaneous music generation and separation},
  author={Mariani, Giorgio and Tallini, Irene and Postolache, Emilian and Mancusi, Michele and Cosmo, Luca and Rodol{\`a}, Emanuele},
  journal={arXiv preprint arXiv:2302.02257},
  year={2023}
}

@inproceedings{iashchenko2023undiff,
  title={UnDiff: Unsupervised Voice Restoration with Unconditional Diffusion Model},
  author={Iashchenko, Anastasiia and Andreev, Pavel and Shchekotov, Ivan and Babaev, Nicholas and Vetrov, Dmitry},
  booktitle={Proc. Interspeech 2023},
  pages={4294--4298},
  year={2023}
}

@article{kong2020diffwave,
  title={Diffwave: A versatile diffusion model for audio synthesis},
  author={Kong, Zhifeng and Ping, Wei and Huang, Jiaji and Zhao, Kexin and Catanzaro, Bryan},
  journal={arXiv preprint arXiv:2009.09761},
  year={2020}
}

@inproceedings{chung2023diffusion,
  title={Diffusion Posterior Sampling for General Noisy Inverse Problems},
  author={Chung, Hyungjin and Kim, Jeongsol and Mccann, Michael T and Klasky, Marc L and Ye, Jong Chul},
  booktitle={The Eleventh International Conference on Learning Representations, ICLR 2023},
  year={2023},
  organization={The International Conference on Learning Representations}
}

@inproceedings{yang2024guidance,
  title={Guidance with spherical gaussian constraint for conditional diffusion},
  author={Yang, Lingxiao and Ding, Shutong and Cai, Yifan and Yu, Jingyi and Wang, Jingya and Shi, Ye},
  booktitle={Proceedings of the 41st International Conference on Machine Learning},
  pages={56071--56095},
  year={2024}
}

@inproceedings{liu2024accelerating,
  title={Accelerating diffusion models for inverse problems through shortcut sampling},
  author={Liu, Gongye and Sun, Haoze and Li, Jiayi and Yin, Fei and Yang, Yujiu},
  booktitle={Proceedings of the Thirty-Third International Joint Conference on Artificial Intelligence},
  pages={1101--1109},
  year={2024}
}

@inproceedings{chung2022come,
  title={Come-closer-diffuse-faster: Accelerating conditional diffusion models for inverse problems through stochastic contraction},
  author={Chung, Hyungjin and Sim, Byeongsu and Ye, Jong Chul},
  booktitle={Proceedings of the IEEE/CVF conference on computer vision and pattern recognition},
  pages={12413--12422},
  year={2022}
}

@inproceedings{moliner2023solving,
  title={Solving audio inverse problems with a diffusion model},
  author={Moliner, Eloi and Lehtinen, Jaakko and V{\"a}lim{\"a}ki, Vesa},
  booktitle={ICASSP 2023-2023 IEEE International Conference on Acoustics, Speech and Signal Processing (ICASSP)},
  pages={1--5},
  year={2023},
  organization={IEEE}
}

@article{xu2025unsupervised,
  title={Unsupervised Blind Speech Separation with a Diffusion Prior},
  author={Xu, Zhongweiyang and Fan, Xulin and Wang, Zhong-Qiu and Jiang, Xilin and Choudhury, Romit Roy},
  journal={arXiv preprint arXiv:2505.05657},
  year={2025}
}

@inproceedings{luo2020dual,
  title={Dual-path rnn: efficient long sequence modeling for time-domain single-channel speech separation},
  author={Luo, Yi and Chen, Zhuo and Yoshioka, Takuya},
  booktitle={ICASSP 2020-2020 IEEE International Conference on Acoustics, Speech and Signal Processing (ICASSP)},
  pages={46--50},
  year={2020},
  organization={IEEE}
}

@article{xu2024tiger,
  title={TIGER: Time-frequency Interleaved Gain Extraction and Reconstruction for Efficient Speech Separation},
  author={Xu, Mohan and Li, Kai and Chen, Guo and Hu, Xiaolin},
  journal={arXiv preprint arXiv:2410.01469},
  year={2024}
}

@article{tian2024u,
  title={U-dits: Downsample tokens in u-shaped diffusion transformers},
  author={Tian, Yuchuan and Tu, Zhijun and Chen, Hanting and Hu, Jie and Xu, Chao and Wang, Yunhe},
  journal={arXiv preprint arXiv:2405.02730},
  year={2024}
}

@article{shazeer2020glu,
  title={Glu variants improve transformer},
  author={Shazeer, Noam},
  journal={arXiv preprint arXiv:2002.05202},
  year={2020}
}

@inproceedings{peebles2023scalable,
  title={Scalable diffusion models with transformers},
  author={Peebles, William and Xie, Saining},
  booktitle={Proceedings of the IEEE/CVF international conference on computer vision},
  pages={4195--4205},
  year={2023}
}

@inproceedings{shi2025distance,
  title={Distance based single-channel target speech extraction},
  author={Shi, Runwu and Yen, Benjamin and Nakadai, Kazuhiro},
  booktitle={ICASSP 2025-2025 IEEE International Conference on Acoustics, Speech and Signal Processing (ICASSP)},
  pages={1--5},
  year={2025},
  organization={IEEE}
}

@article{loshchilov2017decoupled,
  title={Decoupled weight decay regularization},
  author={Loshchilov, Ilya and Hutter, Frank},
  journal={arXiv preprint arXiv:1711.05101},
  year={2017}
}

@article{yu2020gradient,
  title={Gradient surgery for multi-task learning},
  author={Yu, Tianhe and Kumar, Saurabh and Gupta, Abhishek and Levine, Sergey and Hausman, Karol and Finn, Chelsea},
  journal={Advances in neural information processing systems},
  volume={33},
  pages={5824--5836},
  year={2020}
}

@book{makino2018audio,
  title={Audio source separation},
  author={Makino, Shoji},
  volume={433},
  year={2018},
  publisher={Springer}
}

@inproceedings{le2015deep,
  title={Deep NMF for speech separation},
  author={Le Roux, Jonathan and Hershey, John R and Weninger, Felix},
  booktitle={2015 IEEE International Conference on Acoustics, Speech and Signal Processing (ICASSP)},
  pages={66--70},
  year={2015},
  organization={IEEE}
}

@article{wang2014informed,
  title={Informed single-channel speech separation using HMM--GMM user-generated exemplar source},
  author={Wang, Qi and Woo, Wai Lok and Dlay, Satnam Singh},
  journal={IEEE/ACM Transactions on Audio, Speech, and Language Processing},
  volume={22},
  number={12},
  pages={2087--2100},
  year={2014},
  publisher={IEEE}
}

@article{benaroya2005audio,
  title={Audio source separation with a single sensor},
  author={Benaroya, Laurent and Bimbot, Fr{\'e}d{\'e}ric and Gribonval, R{\'e}mi},
  journal={IEEE Transactions on Audio, Speech, and Language Processing},
  volume={14},
  number={1},
  pages={191--199},
  year={2005},
  publisher={IEEE}
}

@article{karamatli2019audio,
  title={Audio source separation using variational autoencoders and weak class supervision},
  author={Karamatl{\i}, Ertu{\u{g}} and Cemgil, Ali Taylan and K{\i}rb{\i}z, Serap},
  journal={IEEE Signal Processing Letters},
  volume={26},
  number={9},
  pages={1349--1353},
  year={2019},
  publisher={IEEE}
}

@article{cao2024survey,
  title={A survey on generative diffusion models},
  author={Cao, Hanqun and Tan, Cheng and Gao, Zhangyang and Xu, Yilun and Chen, Guangyong and Heng, Pheng-Ann and Li, Stan Z},
  journal={IEEE transactions on knowledge and data engineering},
  volume={36},
  number={7},
  pages={2814--2830},
  year={2024},
  publisher={IEEE}
}

@inproceedings{jung2025voicedit,
  title={Voicedit: Dual-condition diffusion transformer for environment-aware speech synthesis},
  author={Jung, Jaemin and Ahn, Junseok and Jung, Chaeyoung and Nguyen, Tan Dat and Jang, Youngjoon and Chung, Joon Son},
  booktitle={ICASSP 2025-2025 IEEE International Conference on Acoustics, Speech and Signal Processing (ICASSP)},
  pages={1--5},
  year={2025},
  organization={IEEE}
}

@inproceedings{ku2025generative,
  title={Generative speech foundation model pretraining for high-quality speech extraction and restoration},
  author={Ku, Pin-Jui and Liu, Alexander H and Korostik, Roman and Huang, Sung-Feng and Fu, Szu-Wei and Juki{\'c}, Ante},
  booktitle={ICASSP 2025-2025 IEEE International Conference on Acoustics, Speech and Signal Processing (ICASSP)},
  pages={1--5},
  year={2025},
  organization={IEEE}
}

@article{luo2019conv,
  title={Conv-tasnet: Surpassing ideal time--frequency magnitude masking for speech separation},
  author={Luo, Yi and Mesgarani, Nima},
  journal={IEEE/ACM transactions on audio, speech, and language processing},
  volume={27},
  number={8},
  pages={1256--1266},
  year={2019},
  publisher={IEEE}
}

@article{wang2023tf,
  title={TF-GridNet: Integrating full-and sub-band modeling for speech separation},
  author={Wang, Zhong-Qiu and Cornell, Samuele and Choi, Shukjae and Lee, Younglo and Kim, Byeong-Yeol and Watanabe, Shinji},
  journal={IEEE/ACM Transactions on Audio, Speech, and Language Processing},
  volume={31},
  pages={3221--3236},
  year={2023},
  publisher={IEEE}
}

@inproceedings{yu2017permutation,
  title={Permutation invariant training of deep models for speaker-independent multi-talker speech separation},
  author={Yu, Dong and Kolb{\ae}k, Morten and Tan, Zheng-Hua and Jensen, Jesper},
  booktitle={2017 IEEE International Conference on Acoustics, Speech and Signal Processing (ICASSP)},
  pages={241--245},
  year={2017},
  organization={IEEE}
}

@article{daras2024survey,
  title={A survey on diffusion models for inverse problems},
  author={Daras, Giannis and Chung, Hyungjin and Lai, Chieh-Hsin and Mitsufuji, Yuki and Ye, Jong Chul and Milanfar, Peyman and Dimakis, Alexandros G and Delbracio, Mauricio},
  journal={arXiv preprint arXiv:2410.00083},
  year={2024}
}

@inproceedings{fei2023generative,
  title={Generative diffusion prior for unified image restoration and enhancement},
  author={Fei, Ben and Lyu, Zhaoyang and Pan, Liang and Zhang, Junzhe and Yang, Weidong and Luo, Tianyue and Zhang, Bo and Dai, Bo},
  booktitle={Proceedings of the IEEE/CVF conference on computer vision and pattern recognition},
  pages={9935--9946},
  year={2023}
}

@inproceedings{saijo2024tf,
  title={TF-Locoformer: Transformer with local modeling by convolution for speech separation and enhancement},
  author={Saijo, Kohei and Wichern, Gordon and Germain, Fran{\c{c}}ois G and Pan, Zexu and Le Roux, Jonathan},
  booktitle={2024 18th International Workshop on Acoustic Signal Enhancement (IWAENC)},
  pages={205--209},
  year={2024},
  organization={IEEE}
}

@article{efron2011tweedie,
  title={Tweedie’s formula and selection bias},
  author={Efron, Bradley},
  journal={Journal of the American Statistical Association},
  volume={106},
  number={496},
  pages={1602--1614},
  year={2011},
  publisher={Taylor \& Francis}
}

@article{song2020score,
  title={Score-based generative modeling through stochastic differential equations},
  author={Song, Yang and Sohl-Dickstein, Jascha and Kingma, Diederik P and Kumar, Abhishek and Ermon, Stefano and Poole, Ben},
  journal={arXiv preprint arXiv:2011.13456},
  year={2020}
}

@article{li2022efficient,
  title={An efficient encoder-decoder architecture with top-down attention for speech separation},
  author={Li, Kai and Yang, Runxuan and Hu, Xiaolin},
  journal={arXiv preprint arXiv:2209.15200},
  year={2022}
}

\cleardoublepage
\appendix

\twocolumn[
  \begin{center}
    {\Large\bfseries Supplementary Material}
  \end{center}
  \vspace{1em}
]

\section{1. Analysis of Separation Process}
\begin{figure}[h]
  \centering
  \includegraphics[width=0.72\columnwidth]{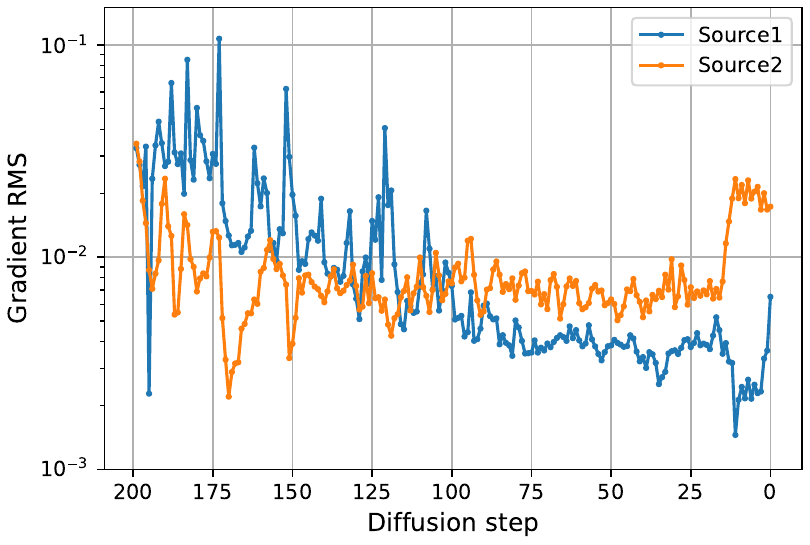}\\
  \caption*{(a) Constant guidance of DPS.}
  \vspace{1em}

  \includegraphics[width=0.72\columnwidth]{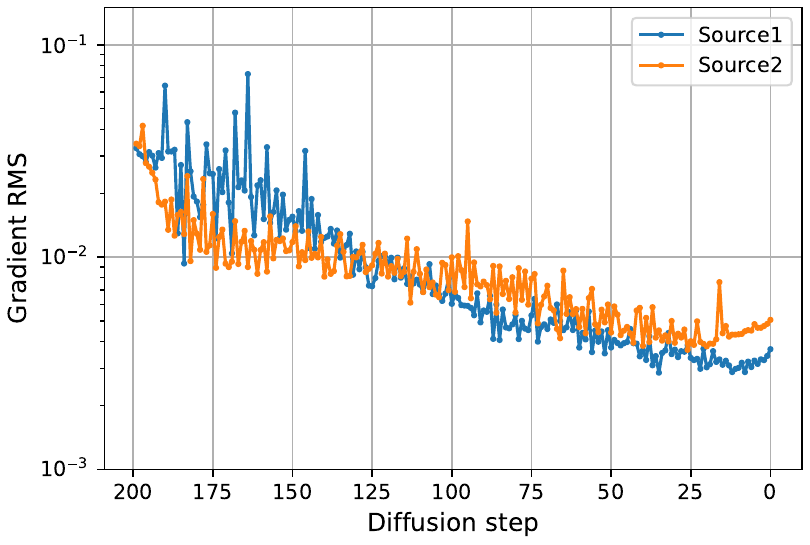}\\
  \caption*{(b) $\sigma(t)$-proportional guidance of DSG.}
  \vspace{1em}

  \includegraphics[width=0.72\columnwidth]{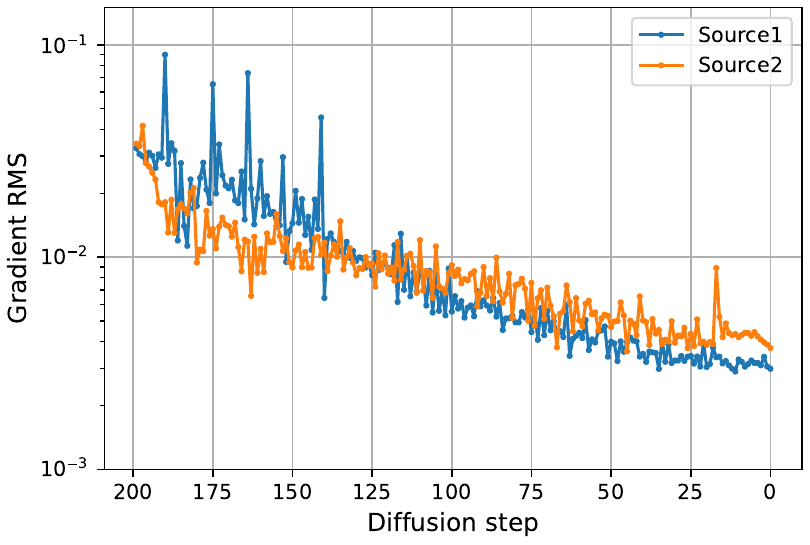}
  \caption*{(c) Our proposed hybrid guidance.}

  \caption{Comparative analysis of gradient RMS evolution.}
  \label{app_fig1}
\end{figure}

We analyze the separation process under different guidance schedules. Detailly, we calculate the Root Mean Square (RMS) of the conditional gradient update using DPS, DSG, and our hybrid guidance schedule $\gamma(t)$, as shown in Figure~\ref{app_fig1}. The constant guidance of DPS shows an unstable gradient update. DSG shows a gradually diminishing gradient update, resulting in a larger update at the end. For comparison, the hybrid $\gamma(t)$ demonstrates more stable separation at the end stage. The separation problem needs non-vanishing guidance because separation demands continuous refinement of sources, especially in the fine-grained, later stages of the diffusion process.

We also observe the intermediate state of sources during the reverse process, from which the estimated $\hat{\bm{x}}_0^k$ can be obtained via Tweedie's formula, providing an effective mechanism to assess whether each source has been "discovered" within the mixture. The following figure illustrates the evolution of $\hat{\bm{x}}_0^k$ for different sources during the reverse process. We present the sum of squares of the estimated $\hat{\bm{x}}_0^k$ as a metric, from which it can be observed that the discovery occurred in the initial stage, where the energy of the estimated clean is unstable, indicating that the source model is discovering the potential source from the noisy intermediate state. 

\begin{figure}[H]
  \centering

  \begin{minipage}{\columnwidth}
    \centering
    \includegraphics[width=0.73\columnwidth]{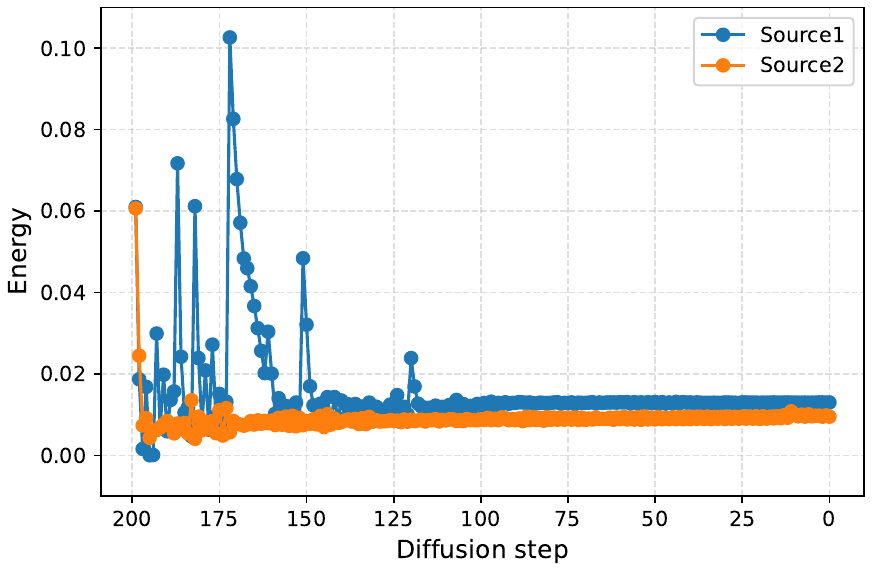}
    \caption*{(a) Constant guidance of DPS.}
  \end{minipage}
  \vspace{1em}

  \begin{minipage}{0.73\columnwidth}
    \centering
    \includegraphics[width=\columnwidth]{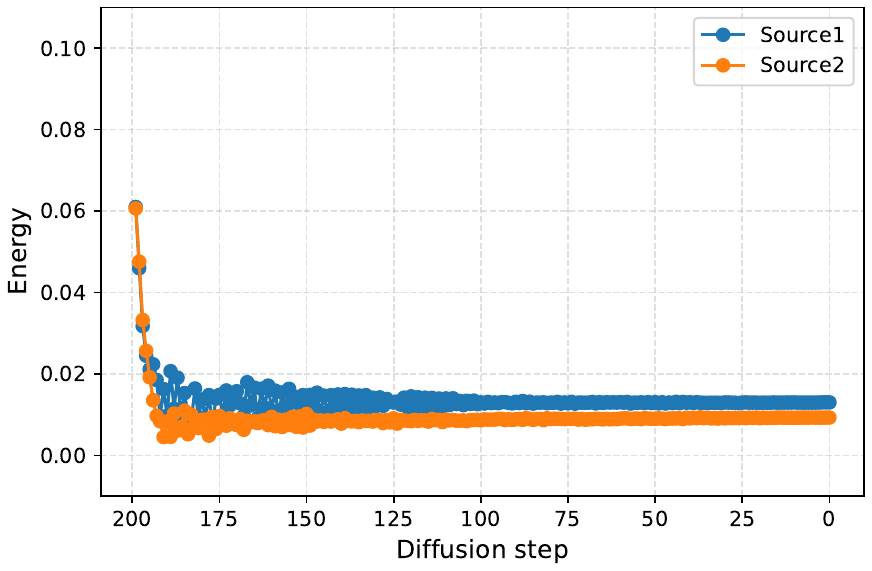}
    \caption*{(b) $\sigma(t)$ proportional guidance of DSG.}
  \end{minipage}
  \vspace{1em}

  \begin{minipage}{0.73\columnwidth}
    \centering
    \includegraphics[width=\columnwidth]{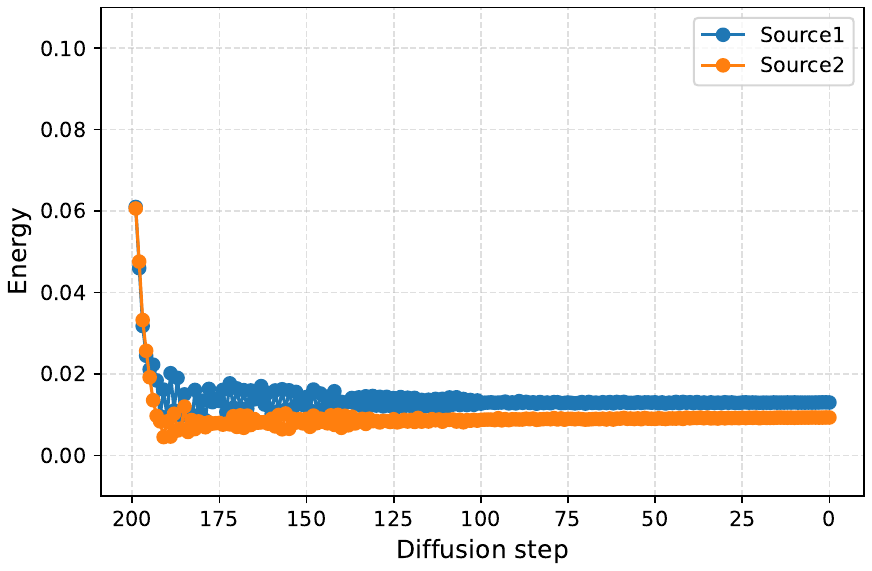}
    \caption*{(c) Our proposed hybrid guidance.}
  \end{minipage}

  \caption{Comparative analysis of estimated $\hat{\bm{x}}_0^k$ evolution.}
  \label{fig2}
\end{figure}

\section{2. Initialization Types}
Our findings indicate that the initialization strategy significantly impacts model performance. In our implementation, we adopt a unified initialization approach where all distinct sources are initialized with identical noisy mixture versions. Additionally, we explore an alternative strategy involving independent source initialization with diverse noise perturbations. To evaluate these approaches, we conduct speech-sound event separation experiments on 50 mixtures while maintaining identical configurations for the guidance schedule $\gamma(t)$ and initialization step $t^*$. Comparative results in Table~\ref{tab:guidance_strategy} demonstrate the performance differences between unified initialization and independent initialization. 

\begin{table}[H] 
\centering
\begin{tabular}{lccc}
\toprule
\textbf{Initialization type} & \textbf{SI-SDR} $\uparrow$ \\
\midrule
Unified Initialization
&
\textbf{14.35}
\\
\addlinespace
Independent Initialization
&
14.19
\\
\bottomrule
\end{tabular}
\caption{Initialization type of different sources.} 
\label{tab:guidance_strategy} 
\end{table}

\section{3. Comparison with Other Models}
To further evaluate the effectiveness of our proposed model, we compare it with representative architectures of similar scale, including a 1D convolutional U-Net baseline and a Diffusion Transformer (DiT) reproduced in the time–frequency (TF) domain. We train both models on the FSD dataset for 150K steps and test the performance on 2 sound event separation on 50 samples. As shown in Table~\ref{tab:compare_models}, our model achieves the highest SI-SDR despite having a comparable number of parameters. Notably, the DiT model, when directly applied in the TF domain, shows degraded separation quality due to its frame-wise processing mechanism, which neglects fine spectral–temporal structures essential for accurate source reconstruction.

\begin{table}[h]
\centering
\begin{tabular}{lccc}
\toprule
\textbf{Model} & \textbf{Params (M)} & \textbf{SI-SDR} $\uparrow$ \\
\midrule
Ours & 37 & \textbf{10.76} \\
1D U-Net & 34 & 6.13 \\
DiT (TF-domain) & 39 & 2.86 \\
\bottomrule
\end{tabular}
\caption{Comparison with other models.}
\label{tab:compare_models}
\end{table}

\section{4. Ablation Study}
We conduct ablation experiments to assess the contribution of each architectural component. The experiment is consistent with the previous section. Instead of simply removing modules, we replace them with functionally neutral alternatives to ensure that parameter counts remain comparable across variants. 
As shown in Table~\ref{tab:if_ablation}, all components provide measurable benefits: removing Intra-Frame Attention leads to the largest degradation, while eliminating Intra-Frequency Attention or Global-Temporal Attention also results in consistent performance drops. 
These results indicate that each module contributes complementary information and jointly supports the overall separation quality.

\begin{table}[h]
\centering
\begin{tabular}{lc}
\toprule
\textbf{Model} & \textbf{SI-SDR} $\uparrow$ \\
\midrule
Full Model & \textbf{10.76} \\
w/o Intra-Frequency Attention & 8.82\\
w/o Intra-Frame Attention & 10.60 \\
w/o Global-Temporal Attention & 10.64 \\
\bottomrule
\end{tabular}
\caption{Ablation study on Intra-Frame Attention, Intra-Frequency Attention, and Global-Temporal Attention.}
\label{tab:if_ablation}
\end{table}

\section{5. Failure Rate Analysis}
We report the failure rate of different separation tasks, as summarized in Table~\ref{tab:failure_rate_transposed}. 
Here, a failure is defined as a case where the averaged SI-SDR of the separated sources falls below 0 dB, indicating that the model fails to generate separated or consistent sources. We report the failure rate on 400 mixtures. The results confirm our method exhibits competitive robustness across most tasks, though the 2 Speech scenario remains the most challenging, with a failure rate of 29.8\%. 
This higher rate primarily arises from instances of speaker inconsistency, where the model may struggle to maintain stable speaker identity during separation. 
In contrast, other tasks such as Speech-Sound separation, 2 Sound separation, and 3 Sound separation demonstrate substantially lower failure rates, confirming that the proposed method performs reliably in multiple source conditions. 
Note that the reported averages in Table~2 include failed cases, ensuring a fair and comprehensive evaluation.

\begin{table}[h]
\small
\begin{tabular}{%
    >{\arraybackslash}p{2.3cm}  %
    >{\centering\arraybackslash}p{1.0cm}    %
    >{\centering\arraybackslash}p{1.5cm}    %
    >{\centering\arraybackslash}p{1.8cm}    %
}
\toprule
\textbf{Tasks} & \textbf{Ours} & \textbf{ConvTasNet} & \textbf{TFLocoformer} \\
\midrule
Speech-Sound     & 1.7\%  & 0.0\%  & 0.0\%  \\
Speech-2 Sound   & 8.8\%  & 4.0\%  & 1.5\%  \\
2 Sound          & 11.5\% & 7.8\%  & 8.3\%  \\
3 Sound          & 23.0\% & 26.5\% & 11.8\% \\
2 Speech         & 29.8\% & 9.3\%  & 2.0\%  \\
\bottomrule
\end{tabular}
\caption{Failure rate (\%) of different separation tasks.}
\label{tab:failure_rate_transposed}
\end{table}

\section{6. Model Complexity and Runtime Efficiency}
We report the model complexity and runtime efficiency for 4-second, 2 sound event separation on an NVIDIA A100 GPU, as summarized in Table~\ref{tab:model_complexity_transposed}. We consider three standard metrics for evaluating computational efficiency: 
(1) \textit{GFLOPs}, which measures the number of floating-point operations required by a single forward pass and reflects the computational complexity of the model; 
(2) \textit{RTF} (Real-Time Factor), defined as the ratio between inference time of the separation algorithm and input audio duration, where higher values indicate slower inference; 
and (3) \textit{GPU memory consumption}, which reports the peak allocated memory during inference. 
These metrics can characterize the computational cost of each model. We measure the computational cost of a single forward pass of each diffusion-based model. Our model achieves a substantially lower GFLOPs requirement compared to Large DiffWave and TF-Locoformer while maintaining a moderate memory footprint.

\begin{table}[h]
\centering
\begin{tabular}{%
    >{\arraybackslash}p{2.1cm}  %
    >{\centering\arraybackslash}p{1.2cm}  %
    >{\centering\arraybackslash}p{1.0cm}  %
    >{\centering\arraybackslash}p{2.3cm}  %
}
\toprule
\textbf{Model}           & \textbf{GFLOPs} & \textbf{RTF} & \textbf{Memory (GiB)} \\
\midrule
Ours               & 394.1           & 12.73       & 10.84            \\
L-DiffWave          & 938.9           & 23.58       & 37.69            \\
ConvTasNet         & 82.9            & 0.02        & 0.54             \\
TFLocoformer      & 1325.8          & 0.04        & 0.13             \\
\bottomrule
\end{tabular}
\caption{Model complexity and computation cost.}
\label{tab:model_complexity_transposed}
\end{table}

\section{7. Model Architecture Details}
Our model contains three kinds of basic blocks for diverse self-attention calculation: Intra-Frame Attention, Intra-Frequency Attention, and Global-Temporal Attention. Intra-Frame Attention and Intra-Frequency Attention blocks, which share the same internal design but differ in attention direction. In detail, the model begins with a convolutional layer that maps the input complex spectrogram $(B, 2, F, T)$ to a $C$-channel feature map $(B, C, F, T)$, which is then sequentially processed by the Intra-Frame and Intra-Frequency attention modules. Specifically, the Intra-Frame Attention reshapes the feature to $(B \cdot T, C, F)$ to apply attention across frequency bins within each time frame, while the Intra-Frequency Attention reshapes it to $(B \cdot F, C, T)$ to focus on the temporal evolution within each frequency bin. This strategy has been adopted in various speech processing models~\cite{luo2020dual, xu2024tiger}, allowing the attention mechanism to capture fine-grained spectral and temporal features. The Global-Temporal Attention block is designed to capture long-range temporal dependencies across the entire frequency band. To reduce the computational cost, the input feature map $(B, C, F, T)$ is first unshuffled along the frequency axis, effectively downsampling the frequency resolution and producing a tensor of shape $(B, C \cdot {N}_{F}, F/{N}_{F}, T)$. A SwiGLU module is then applied to project the channel dimension to $(B, C', F/{N}_{F}, T)$. The feature map is subsequently reshaped to $(B, C' \cdot F/{N}_{F}, T)$ for attention computation across the full-time axis. After this, an inverse operation restores the feature map to its original shape. This block is exclusively used in the third latent layer of the U-Net. The model configuration is presented in Table~\ref{tab:model}. 

\begin{table}[H] 
\centering
\small 
\begin{tabular}{
    >{\centering\arraybackslash}m{3cm}
    >{\centering\arraybackslash}m{3.2cm}
}
\toprule
\textbf{Configuration} & \textbf{Parameter} \\
\midrule
$F$
&
256
\\
\addlinespace
$T$
&
252 (4s, 16kHz)
\\
\addlinespace
$C$
&
72
\\
\addlinespace

$C'$
&
16
\\
\addlinespace
$N_F$
&
4

\\

\bottomrule
\end{tabular}
\caption{Model configuration.} 
\label{tab:model} 
\end{table}

\section{8. Experiment Details}
Conv-TasNet \cite{luo2019conv} and TF-Locoformer \cite{saijo2024tf} are adopted as supervised baselines. Conv-TasNet is a classical method proposed for speech separation in the time domain. TF-Locoformer is an advanced TF domain method in both speech separation and universal audio separation tasks. For model training, we randomly generate 2,000 mixtures and train each model on each task for 200 epochs using PIT training with SI-SDR as the loss function. During training, each sources have a random temporal offset.

\begin{table}[H] 
\small 
\begin{tabular}{
    >{\arraybackslash}m{6cm}
    >{\centering\arraybackslash}m{1.5cm}
}
\toprule
\textbf{Configuration} & \textbf{Parameter} \\
\midrule
Number of filters in autoencoder
&
$N=512$
\\
\addlinespace
Length of the filters
&
$L=16$
\\
\addlinespace
Number of channels in bottleneck
&
$B=128$
\\
\addlinespace
Number of channels in convolutional blocks
&
$H=512$
\\
\addlinespace
Number of channels in skip-connection
&
$Sc=128$
\\

\addlinespace
Kernel size
&
$P=3$
\\
\addlinespace
Stacked separation blocks
&
$X=8$
\\
\addlinespace
Repetitions per block
&
$R=3$
\\
\bottomrule
\end{tabular}
\caption{Conv-TasNet's configuration.} 
\label{tab:conv} 
\end{table}

For TF-Locoformer, we adopt the official implementation\footnote{\url{https://github.com/merlresearch/tf-locoformer}}, and adopt the small model with the parameter of 5.0M. The configuration is presented in Table~\ref{tab:tf}.

\begin{table}[H] 
\centering
\small 
\begin{tabular}{
    >{\arraybackslash}m{6cm}
    >{\centering\arraybackslash}m{1.5cm}
}
\toprule
\textbf{Configuration} & \textbf{Parameter} \\
\midrule
Embedding dimension of each TF bin
&
$D=96$
\\
\addlinespace
Number of Locoformer blocks
&
$B=4$
\\
\addlinespace
Hidden dimension in Conv-SwiGLU
&
$C=256$
\\
\addlinespace
Kernel size in Conv1D and Deconv1D
&
$K=4$
\\
\addlinespace
Stride in Conv1D and Deconv1D
&
$S=1$
\\
\addlinespace
Number of heads in self-attention
&
$H=4$
\\
\addlinespace
Number of groups in RMSGroupNorm
&
$G=4$
\\

\bottomrule
\end{tabular}
\caption{TF-Locoformer's configuration.} 
\label{tab:tf} 
\end{table}

\cleardoublepage
\twocolumn[
  \begin{center}
    \section{9. Result Visualization}
    \subsection*{Speech-Sound event separation}
  \end{center}
  \vspace{1em}
]


\begin{figure}[H]
\centering
\includegraphics[width=0.23\textwidth]{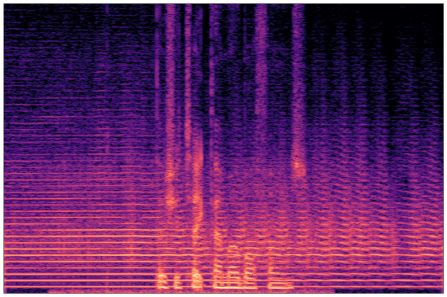} %
\caption{Mixture consisting of a speech and a sound event.}
\label{fig13}
\end{figure}
\begin{figure}[H]
\centering
\begin{minipage}{0.23\textwidth}
    \centering
    \includegraphics[width=\textwidth]{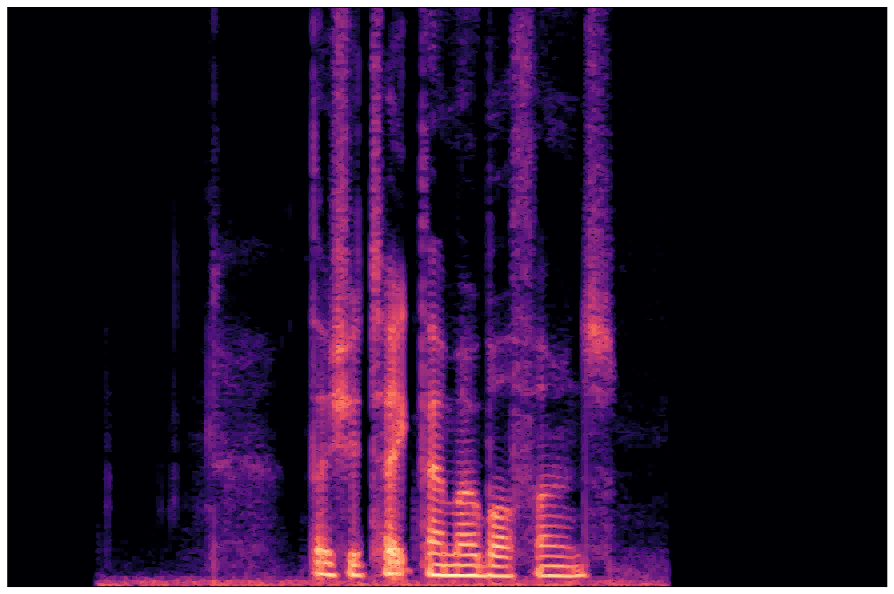}
\end{minipage}
\begin{minipage}{0.23\textwidth}
    \centering
    \includegraphics[width=\textwidth]{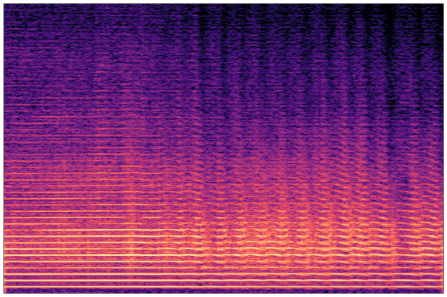} %
\end{minipage}
\caption{Ground truth of sources.}
\label{fig14}
\end{figure}
\begin{figure}[H]
\centering
\begin{minipage}{0.23\textwidth}
    \centering
    \includegraphics[width=\textwidth]{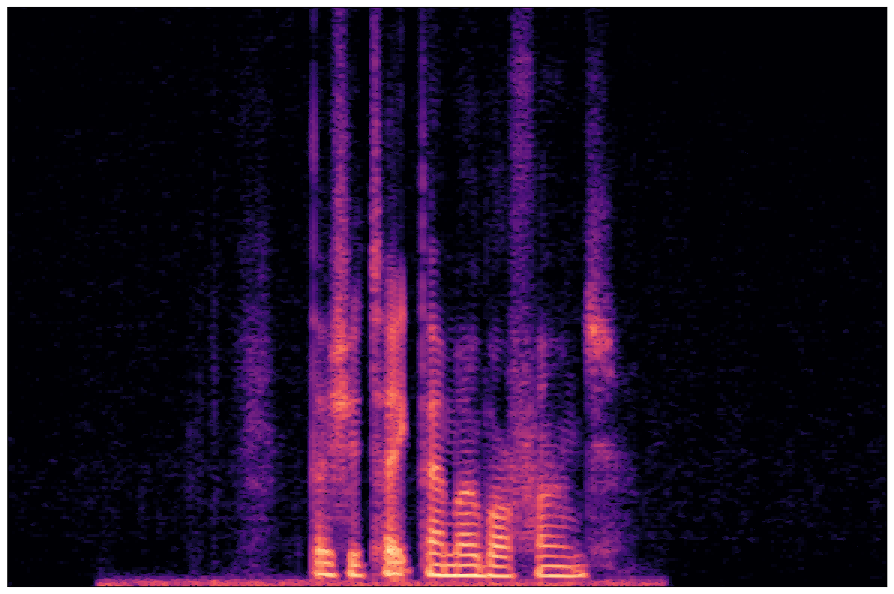}
\end{minipage}
\begin{minipage}{0.23\textwidth}
    \centering
    \includegraphics[width=\textwidth]{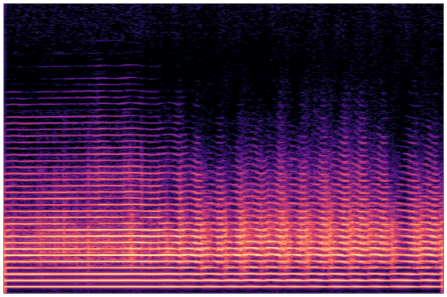} 
\end{minipage}
\caption{Separation of our method.}
\label{fig15}
\end{figure}
\begin{figure}[H]
\centering
\begin{minipage}{0.23\textwidth}
    \centering
    \includegraphics[width=\textwidth]{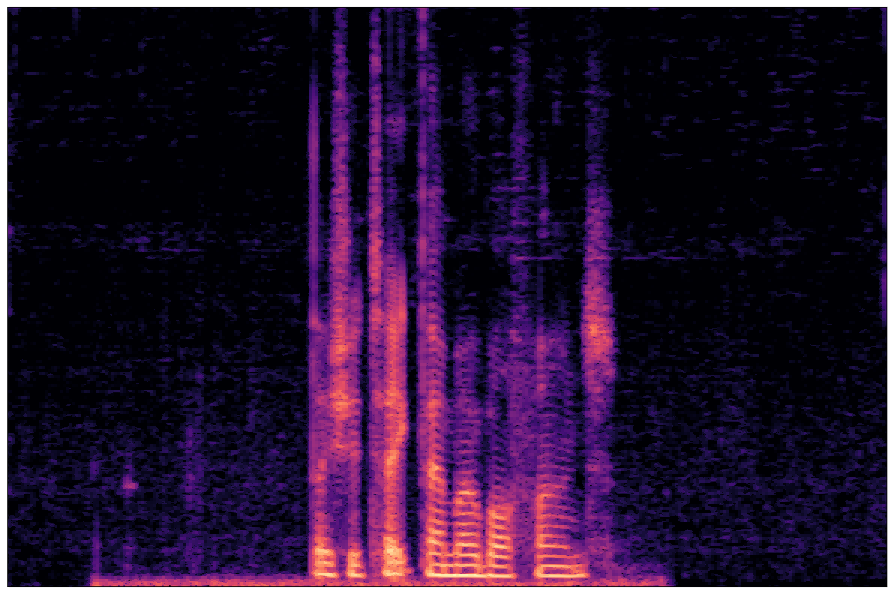}
\end{minipage}
\begin{minipage}{0.23\textwidth}
    \centering
    \includegraphics[width=\textwidth]{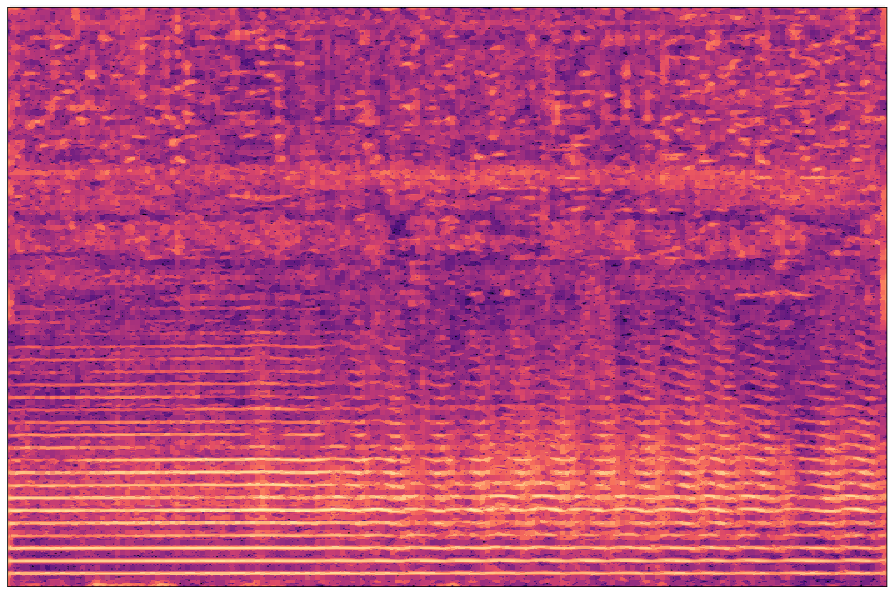} 
\end{minipage}
\caption{Separation of constant guidance.}
\label{fig16}
\end{figure}
\begin{figure}[H]
\centering
\begin{minipage}{0.23\textwidth}
    \centering
    \includegraphics[width=\textwidth]{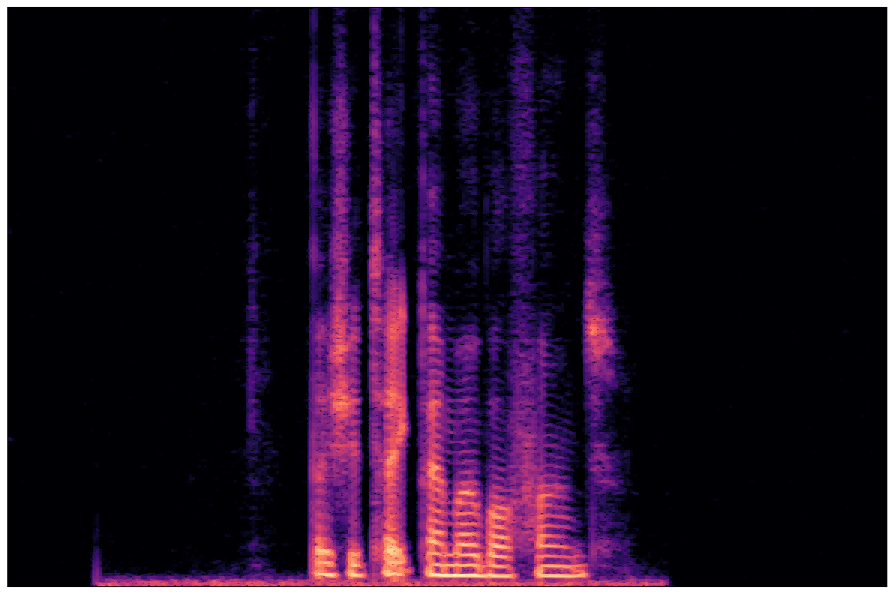}
\end{minipage}
\begin{minipage}{0.23\textwidth}
    \centering
    \includegraphics[width=\textwidth]{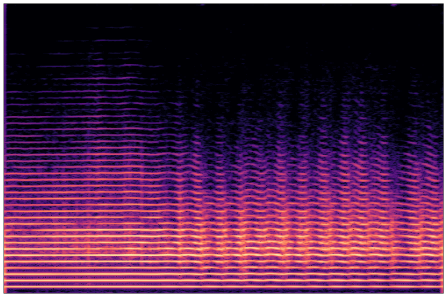} 
\end{minipage}
\caption{Separation of $\sigma(t)$ proportional guidance.}
\label{fig17}
\end{figure}
\begin{figure}[H]
\centering
\begin{minipage}{0.23\textwidth}
    \centering
    \includegraphics[width=\textwidth]{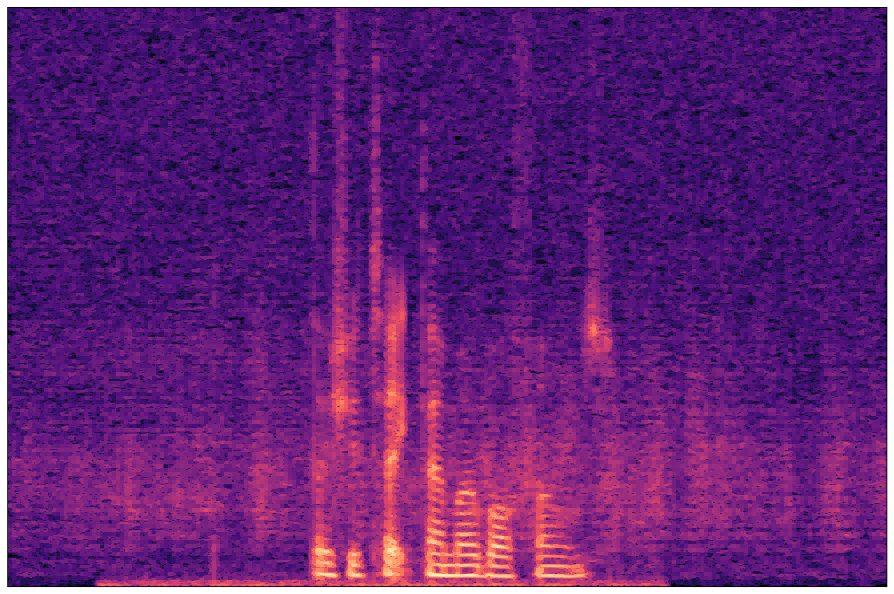}
\end{minipage}
\begin{minipage}{0.23\textwidth}
    \centering
    \includegraphics[width=\textwidth]{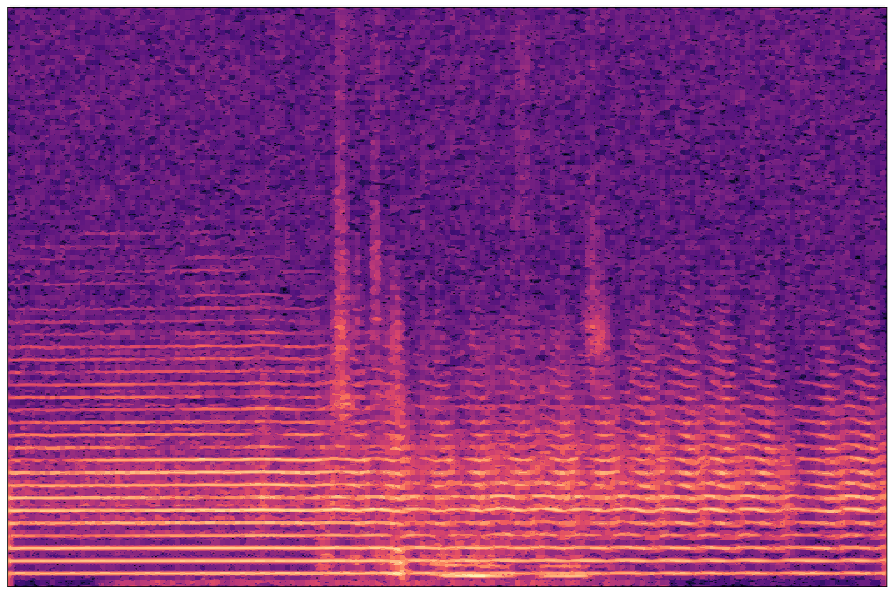} 
\end{minipage}
\caption{Separation of Analytic sampling.}
\label{fig18}
\end{figure}
\begin{figure}[H]
\centering
\begin{minipage}{0.23\textwidth}
    \centering
    \includegraphics[width=\textwidth]{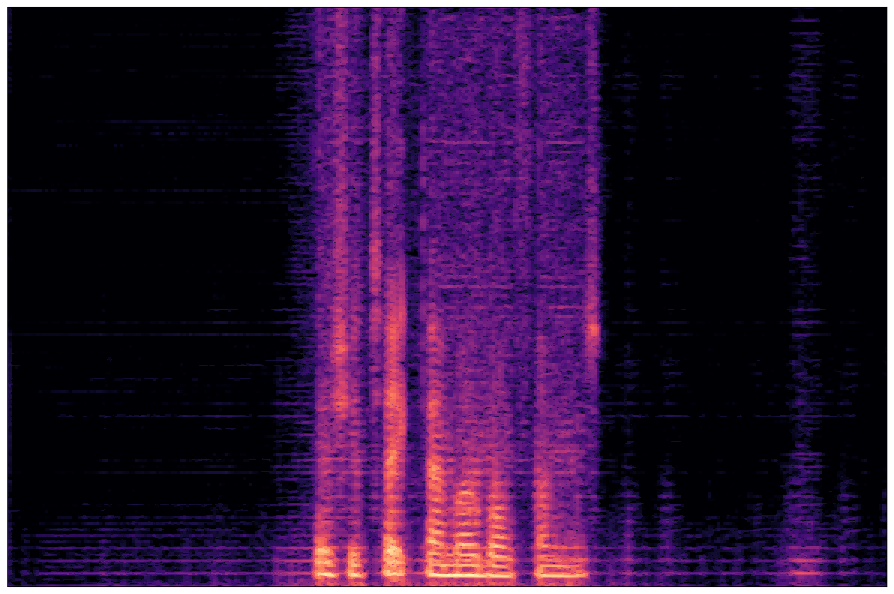}
\end{minipage}
\begin{minipage}{0.23\textwidth}
    \centering
    \includegraphics[width=\textwidth]{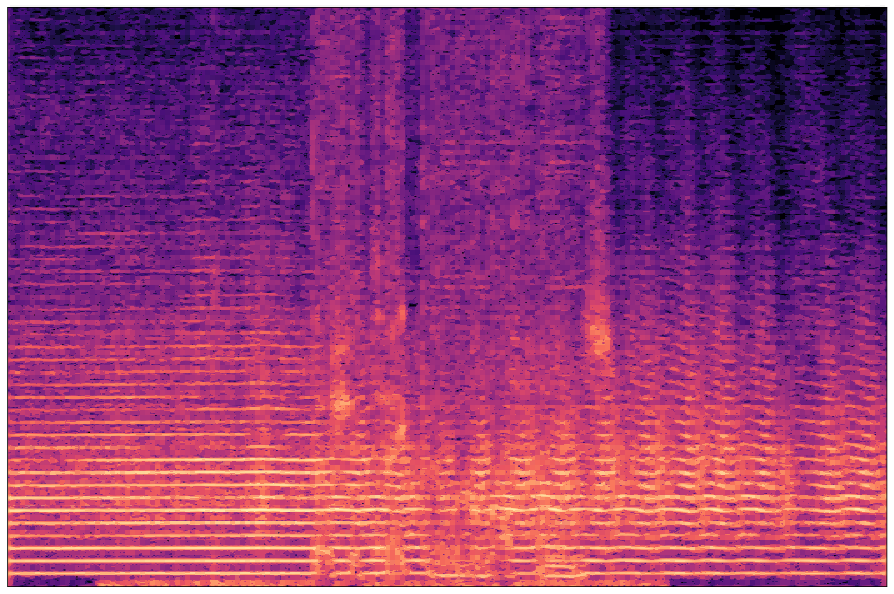} 
\end{minipage}
\caption{Separation of Conv-TasNet.}
\label{fig19}
\end{figure}
\begin{figure}[H]
\centering
\begin{minipage}{0.23\textwidth}
    \centering
    \includegraphics[width=\textwidth]{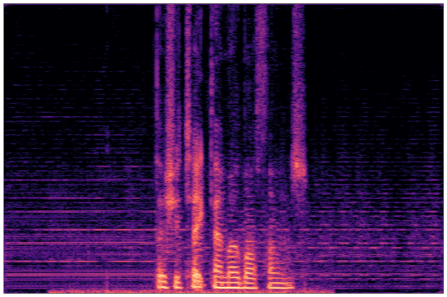}
\end{minipage}
\begin{minipage}{0.23\textwidth}
    \centering
    \includegraphics[width=\textwidth]{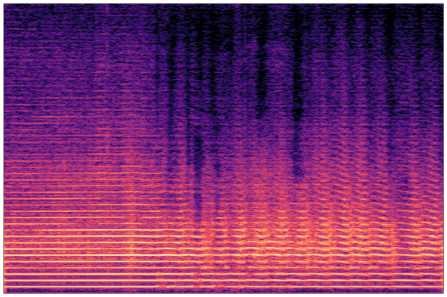} 
\end{minipage}
\caption{Separation of TF-Locoformer.}
\label{fig20}
\end{figure}

\clearpage
\twocolumn[
  \begin{center}
    \subsection*{Sound Event separation}
  \end{center}
  \vspace{1em}
]

\begin{figure}[H]
\centering
\includegraphics[width=0.23\textwidth]{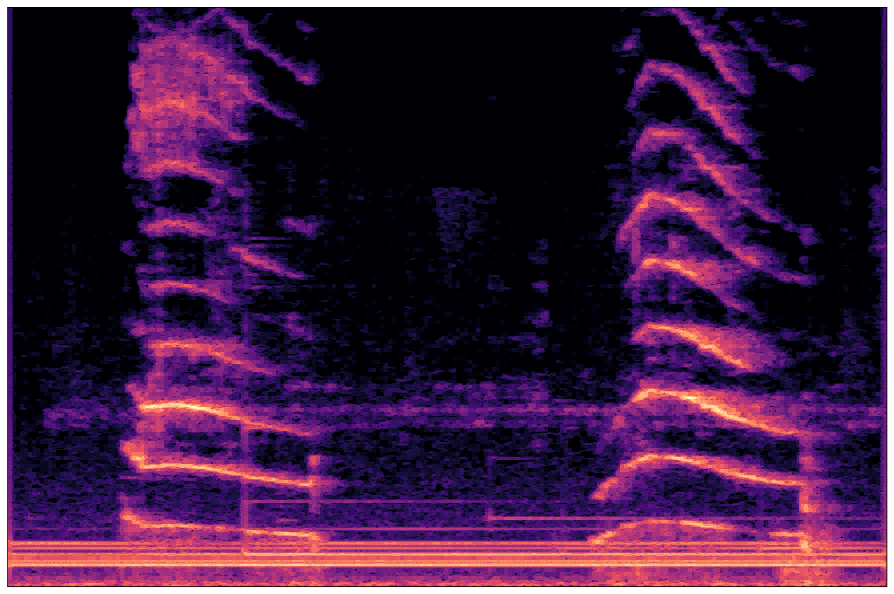} %
\caption{Mixture consisting of an instrument and meow.}
\label{fig3}
\end{figure}

\begin{figure}[H]
\centering
\begin{minipage}{0.23\textwidth}
    \centering
    \includegraphics[width=\textwidth]{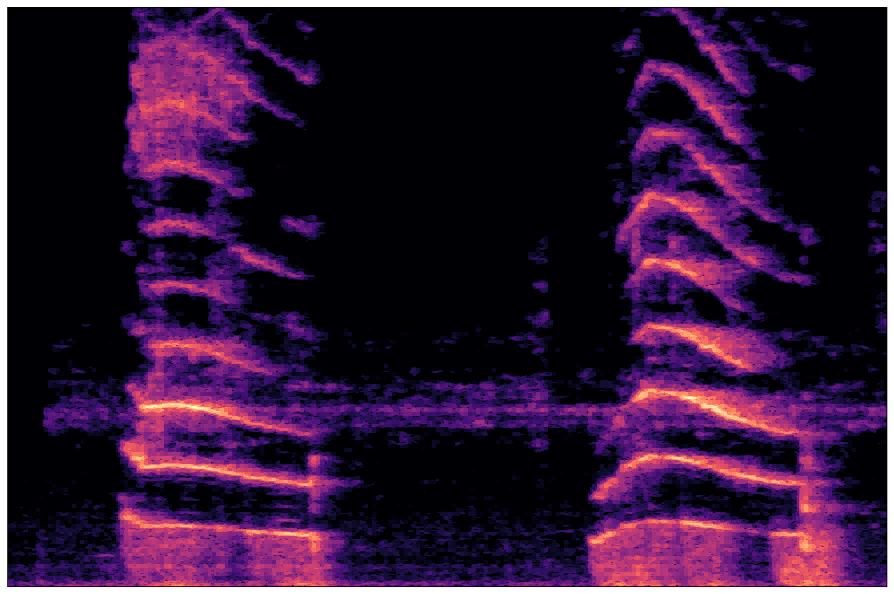}
\end{minipage}
\begin{minipage}{0.23\textwidth}
    \centering
    \includegraphics[width=\textwidth]{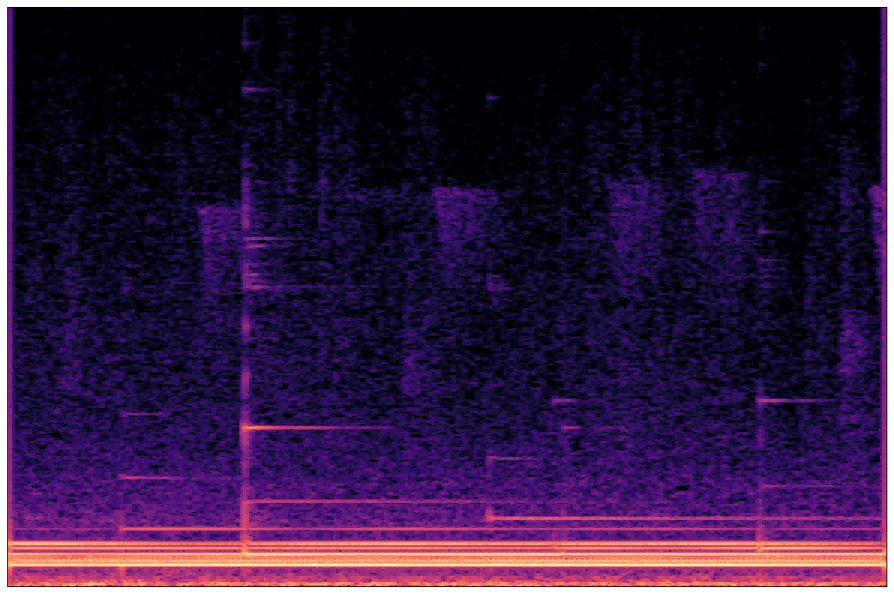} 
\end{minipage}
\caption{Ground truth of sources.}
\label{fig4}
\end{figure}

\begin{figure}[H]
\centering
\begin{minipage}{0.23\textwidth}
    \centering
    \includegraphics[width=\textwidth]{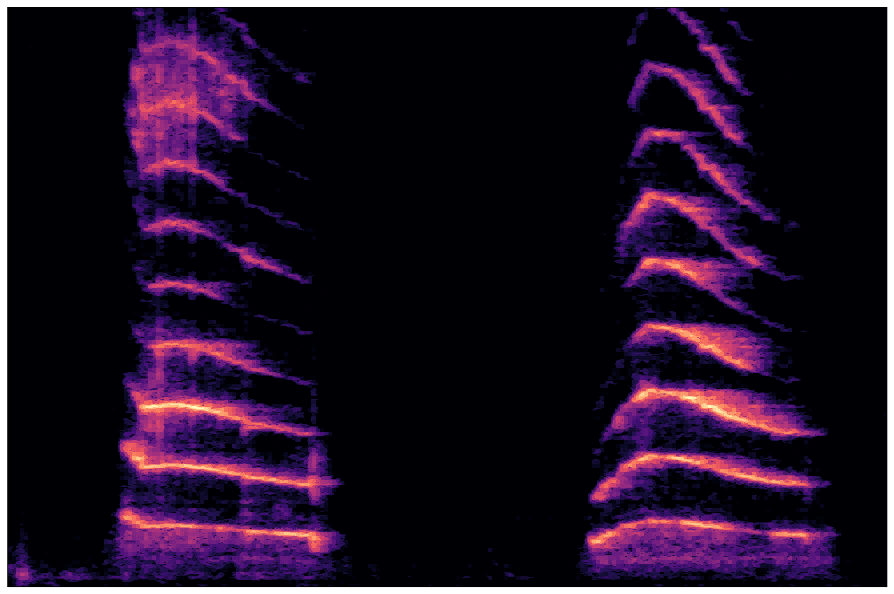}
\end{minipage}
\begin{minipage}{0.23\textwidth}
    \centering
    \includegraphics[width=\textwidth]{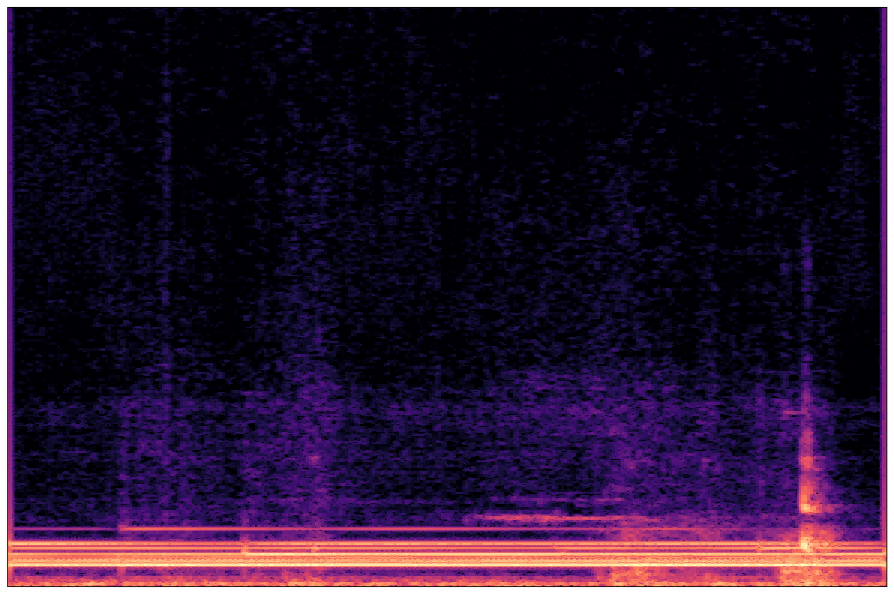} 
\end{minipage}
\caption{Separation of our method.}
\label{fig5}
\end{figure}

\begin{figure}[H]
\centering
\begin{minipage}{0.23\textwidth}
    \centering
    \includegraphics[width=\textwidth]{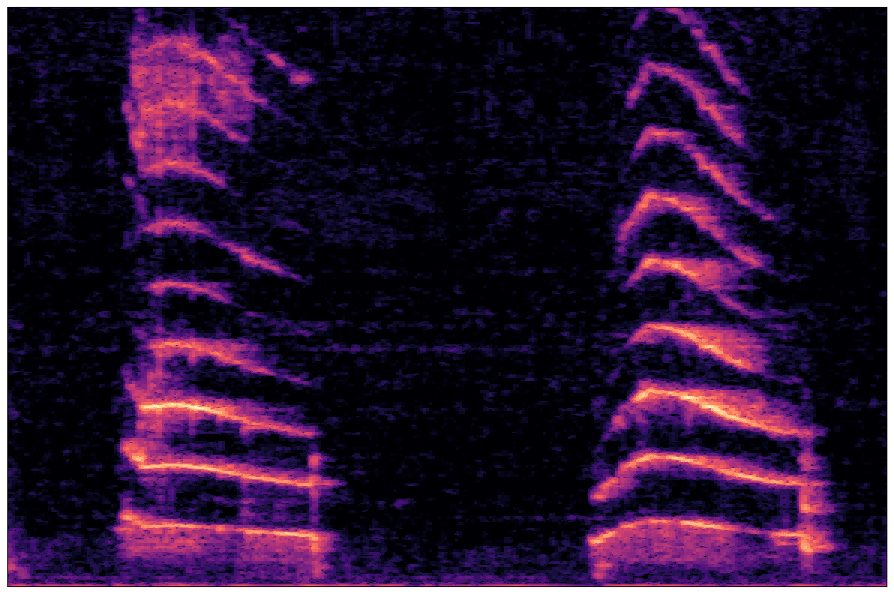}
\end{minipage}
\begin{minipage}{0.23\textwidth}
    \centering
    \includegraphics[width=\textwidth]{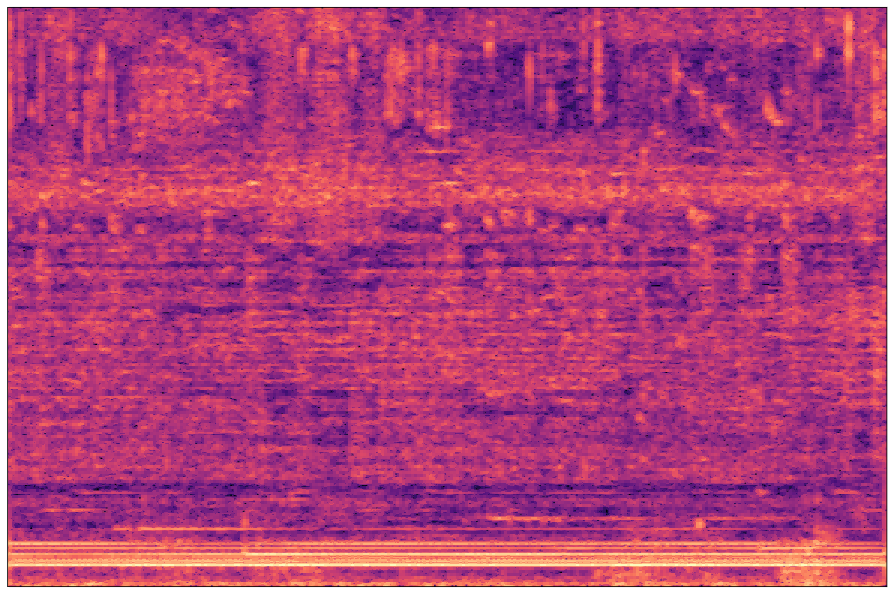} 
\end{minipage}
\caption{Separation of constant guidance.}
\label{fig6}
\end{figure}

\begin{figure}[H]
\centering
\begin{minipage}{0.23\textwidth}
    \centering
    \includegraphics[width=\textwidth]{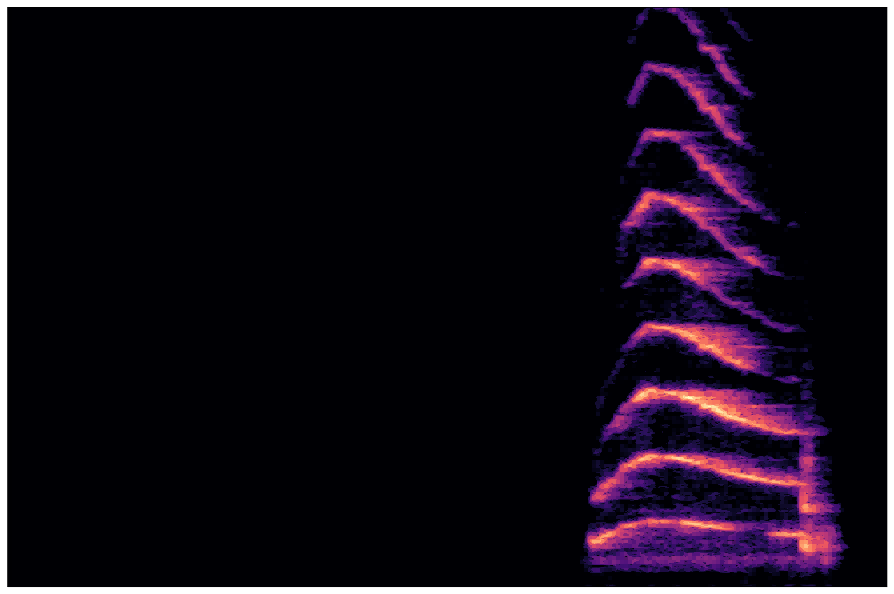}
\end{minipage}
\begin{minipage}{0.23\textwidth}
    \centering
    \includegraphics[width=\textwidth]{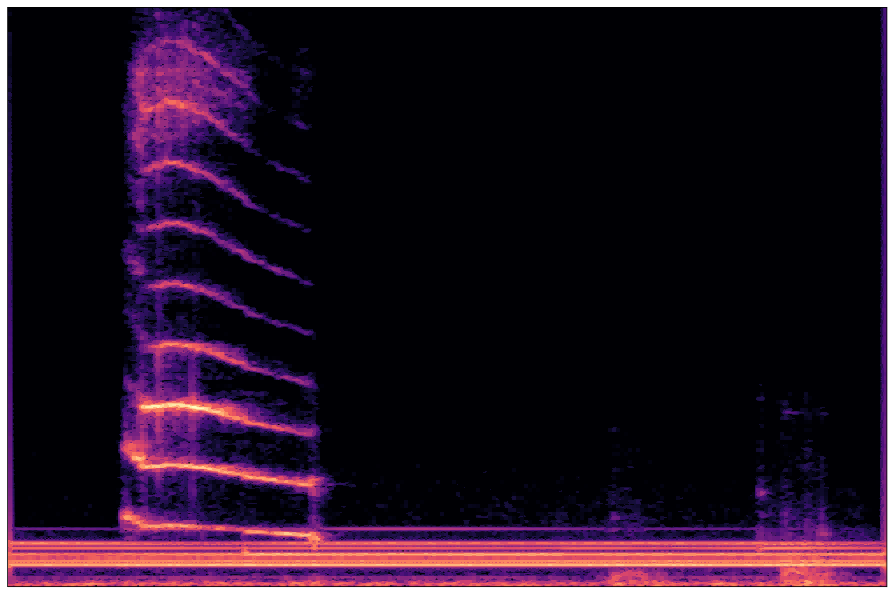} 
\end{minipage}
\caption{Separation of $\sigma(t)$ proportional guidance.}
\label{fig7}
\end{figure}

\begin{figure}[H]
\centering
\begin{minipage}{0.23\textwidth}
    \centering
    \includegraphics[width=\textwidth]{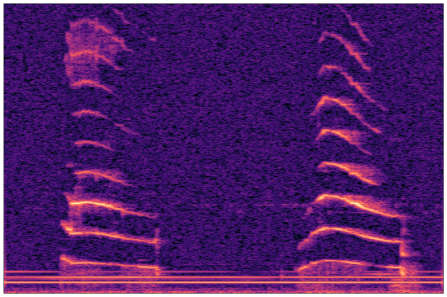}
\end{minipage}
\begin{minipage}{0.23\textwidth}
    \centering
    \includegraphics[width=\textwidth]{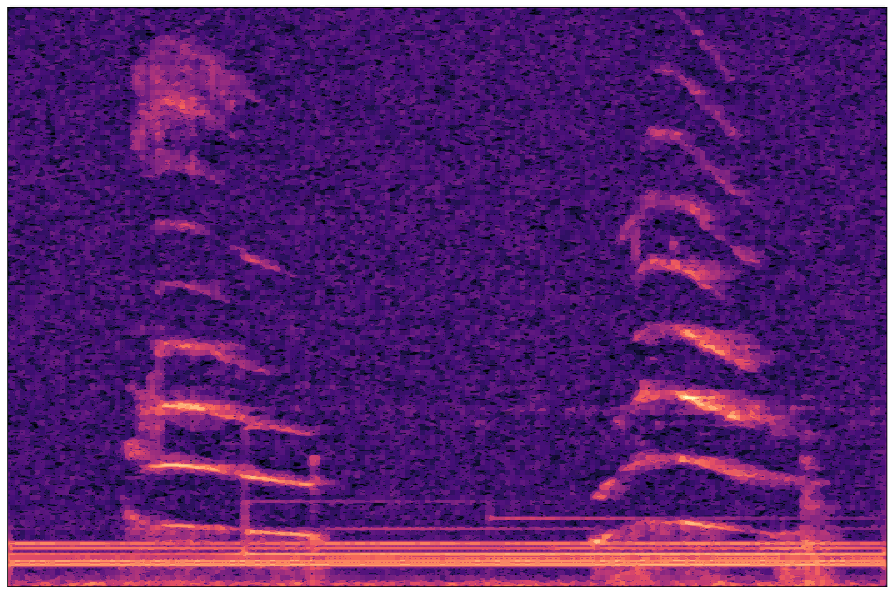} 
\end{minipage}
\caption{Separation of Analytic sampling.}
\label{fig8}
\end{figure}

\begin{figure}[H]
\centering
\begin{minipage}{0.23\textwidth}
    \centering
    \includegraphics[width=\textwidth]{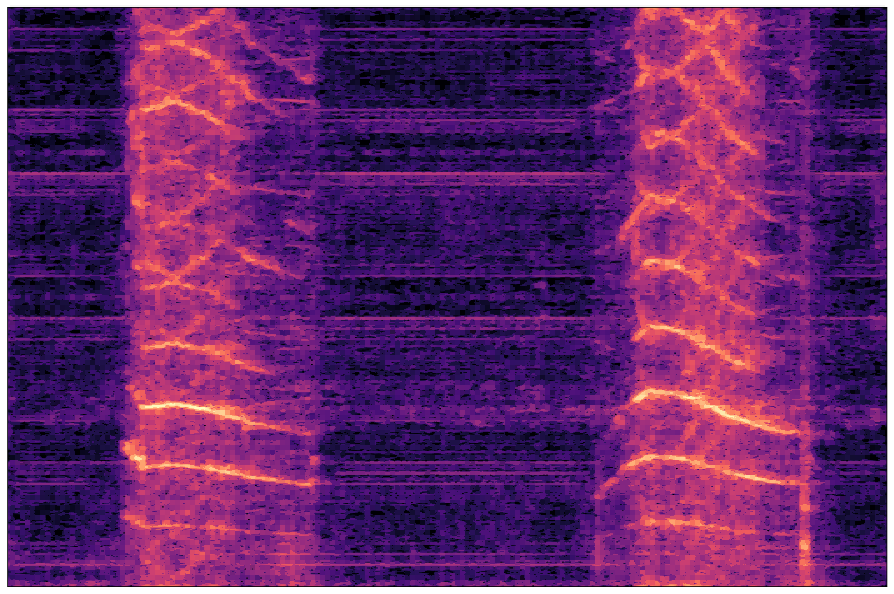}
\end{minipage}
\begin{minipage}{0.23\textwidth}
    \centering
    \includegraphics[width=\textwidth]{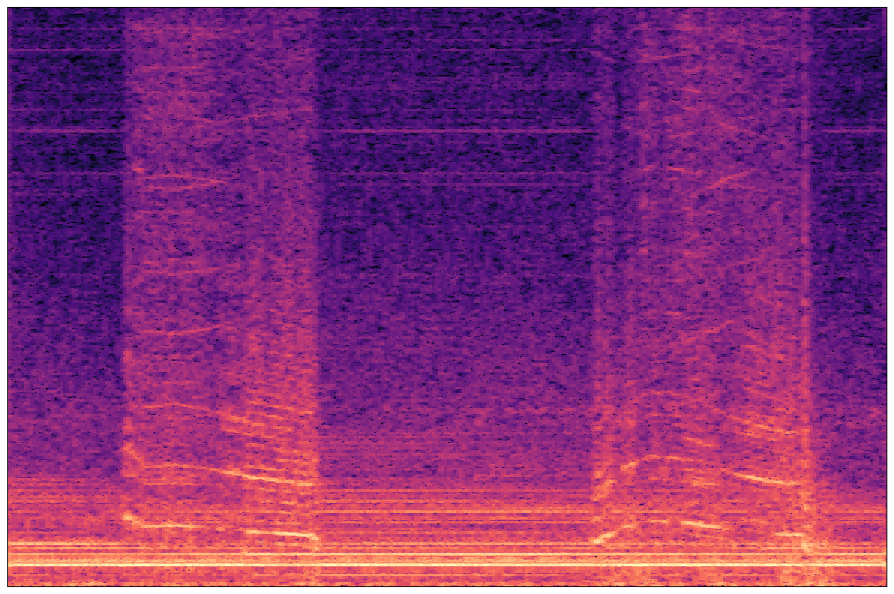} 
\end{minipage}
\caption{Separation of Conv-TasNet.}
\label{fig9}
\end{figure}

\begin{figure}[H]
\centering
\begin{minipage}{0.23\textwidth}
    \centering
    \includegraphics[width=\textwidth]{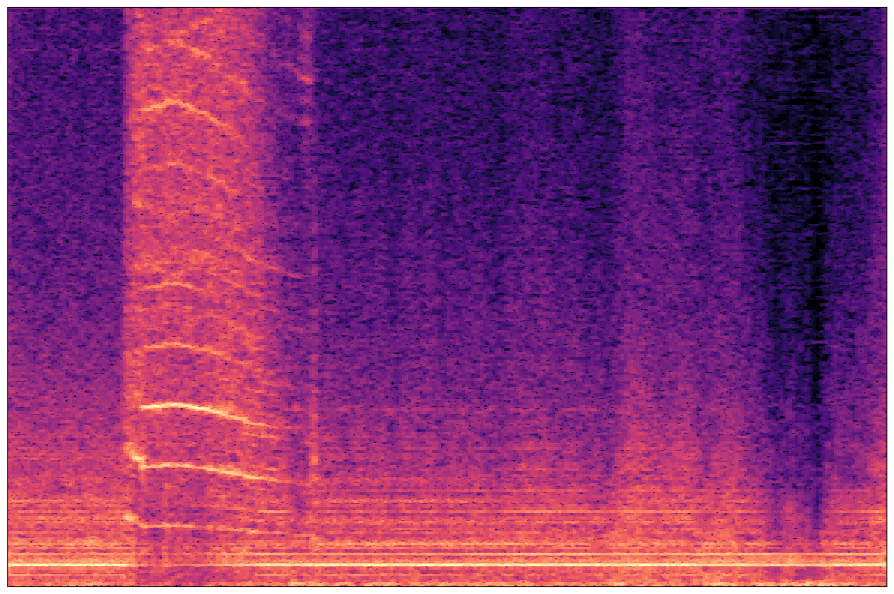}
\end{minipage}
\begin{minipage}{0.23\textwidth}
    \centering
    \includegraphics[width=\textwidth]{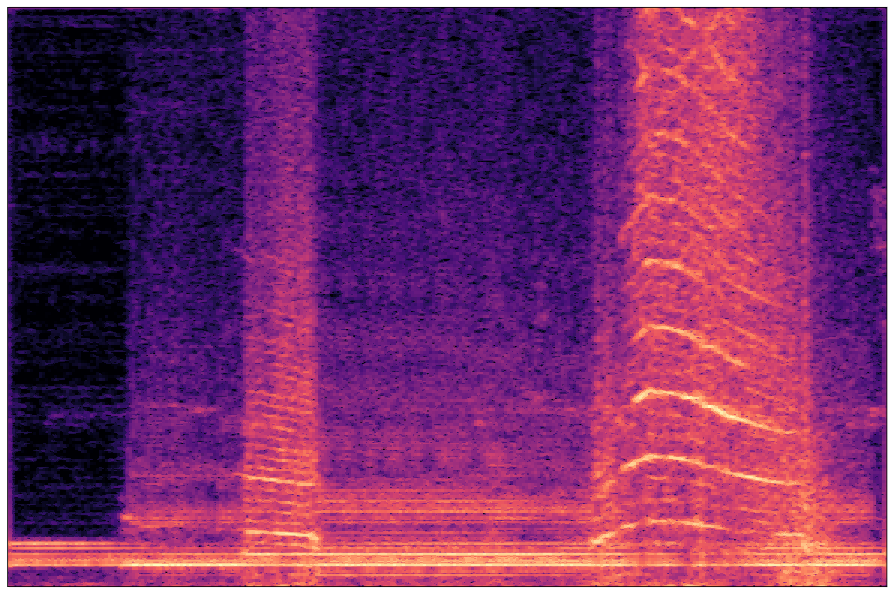} 
\end{minipage}
\caption{Separation of TF-Locoformer.}
\label{fig10}
\end{figure}

\clearpage

\twocolumn[
  \begin{center}
    \subsection*{Speech separation}
  \end{center}
  \vspace{1em}
]


\begin{figure}[H]
\centering
\includegraphics[width=0.23\textwidth]{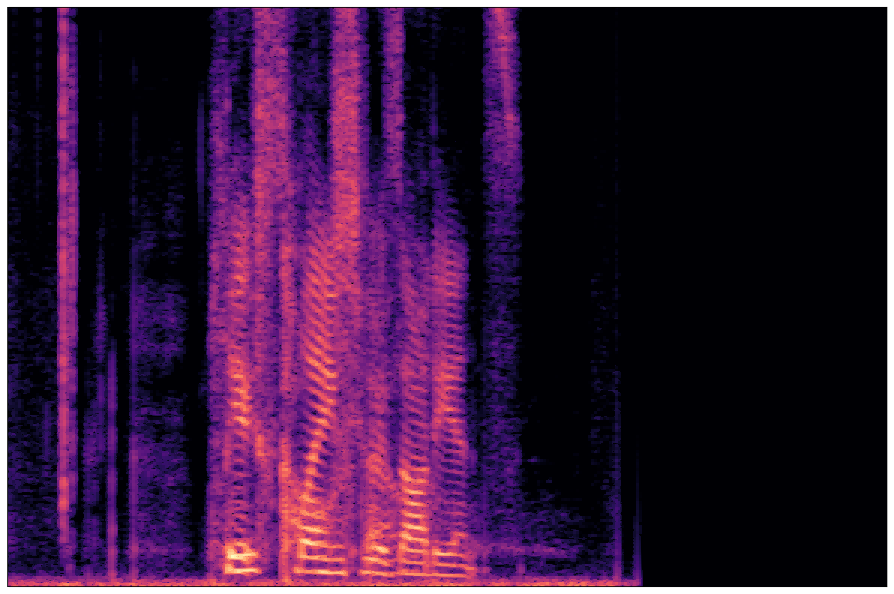} %
\caption{Mixture consisting of two speeches.}
\label{fig21}
\end{figure}

\begin{figure}[H]
\centering
\begin{minipage}{0.23\textwidth}
    \centering
    \includegraphics[width=\textwidth]{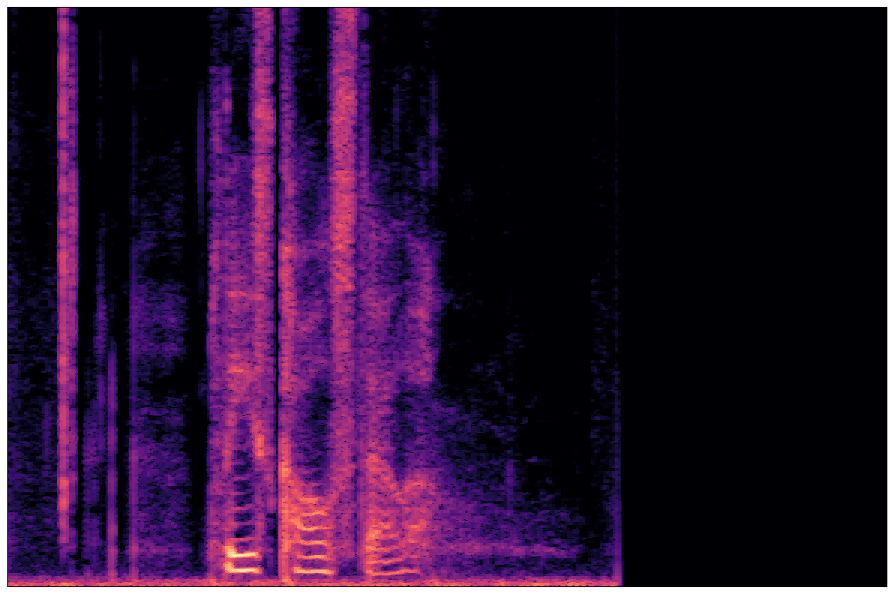}
\end{minipage}
\begin{minipage}{0.23\textwidth}
    \centering
    \includegraphics[width=\textwidth]{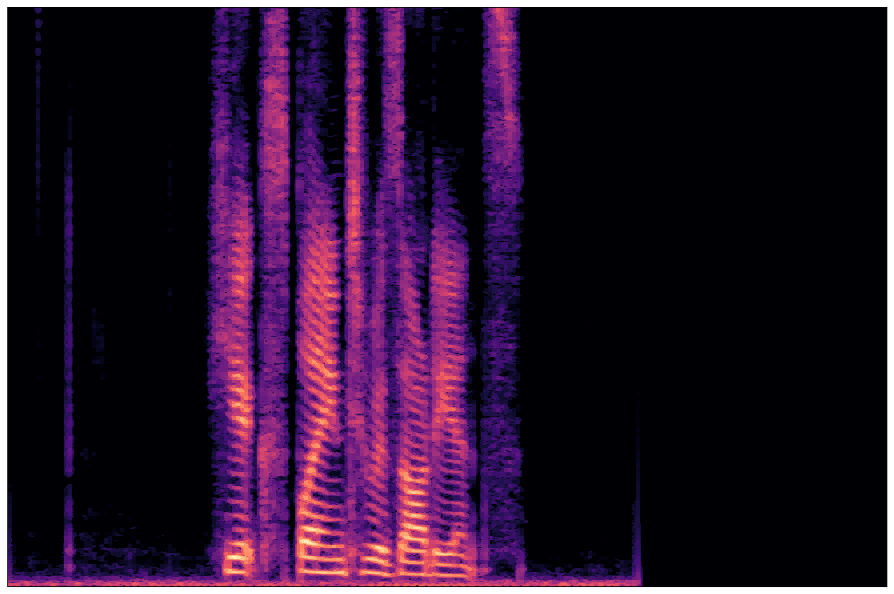} 
\end{minipage}
\caption{Ground truth of sources.}
\label{fig22}
\end{figure}

\begin{figure}[H]
\centering
\begin{minipage}{0.23\textwidth}
    \centering
    \includegraphics[width=\textwidth]{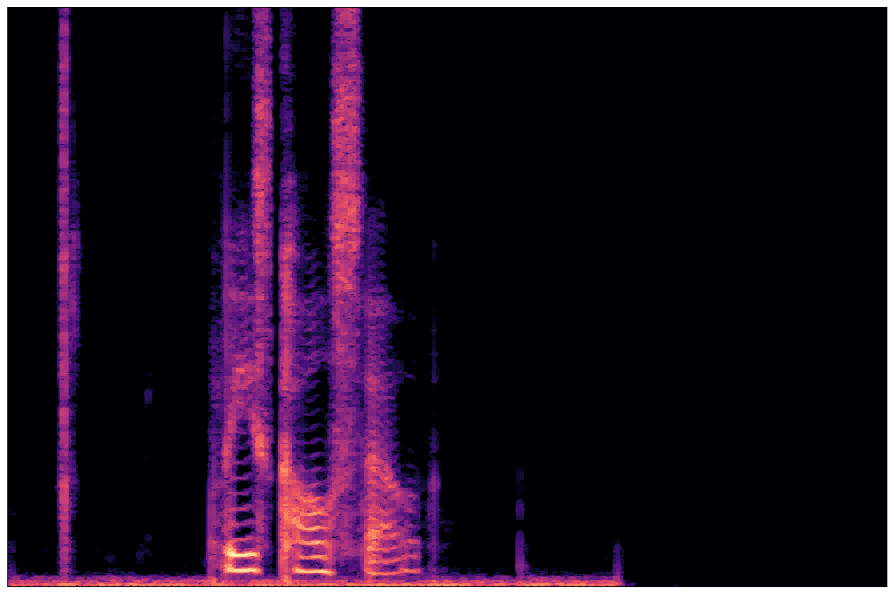}
\end{minipage}
\begin{minipage}{0.23\textwidth}
    \centering
    \includegraphics[width=\textwidth]{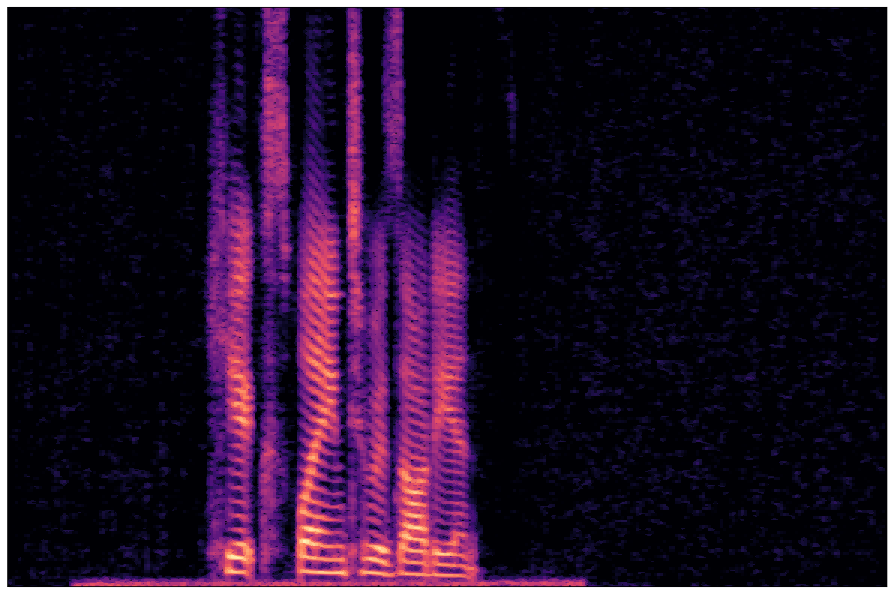} 
\end{minipage}
\caption{Separation of our method using our model.}
\label{fig23}
\end{figure}

\begin{figure}[H]
\centering
\begin{minipage}{0.23\textwidth}
    \centering
    \includegraphics[width=\textwidth]{figs3/hybrid/sepa_0.png}
\end{minipage}
\begin{minipage}{0.23\textwidth}
    \centering
    \includegraphics[width=\textwidth]{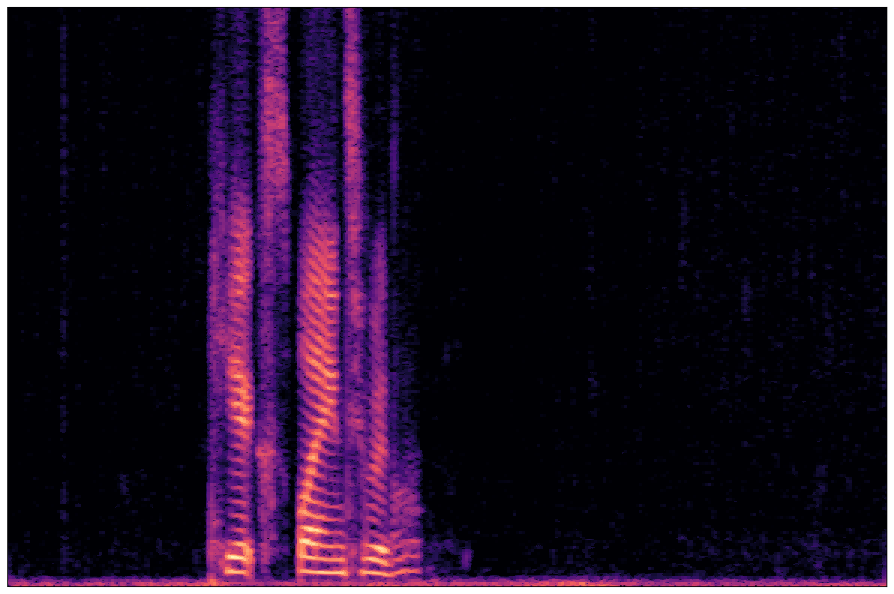} 
\end{minipage}
\caption{Separation of our method using Large Diffwave.}
\label{fig29}
\end{figure}

\begin{figure}[H]
\centering
\begin{minipage}{0.23\textwidth}
    \centering
    \includegraphics[width=\textwidth]{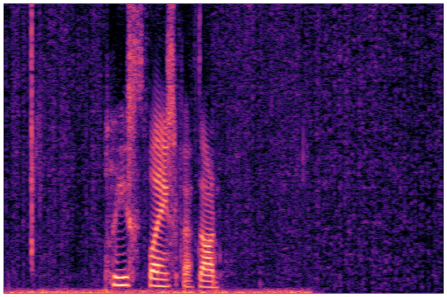}
\end{minipage}
\begin{minipage}{0.23\textwidth}
    \centering
    \includegraphics[width=\textwidth]{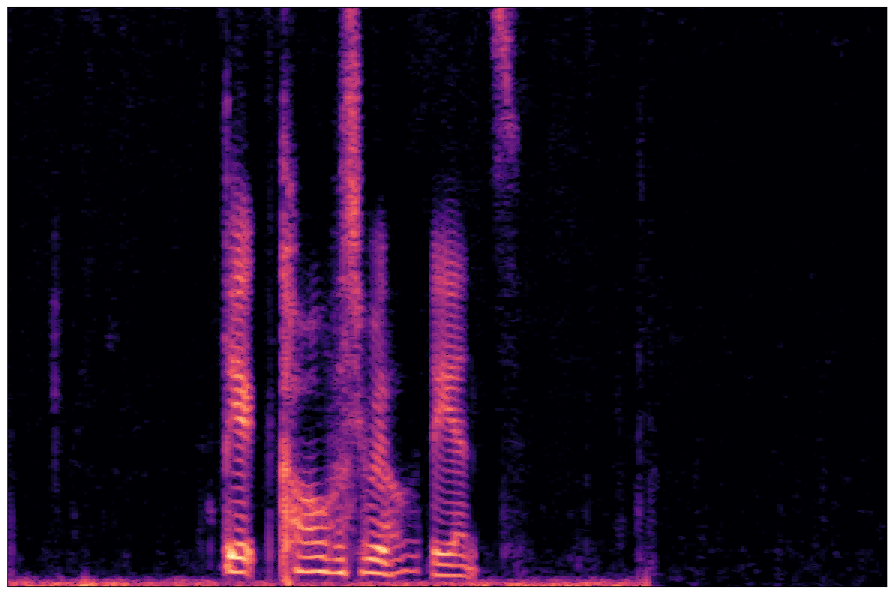} 
\end{minipage}
\caption{Separation of constant guidance.}
\label{fig24}
\end{figure}

\begin{figure}[H]
\centering
\begin{minipage}{0.23\textwidth}
    \centering
    \includegraphics[width=\textwidth]{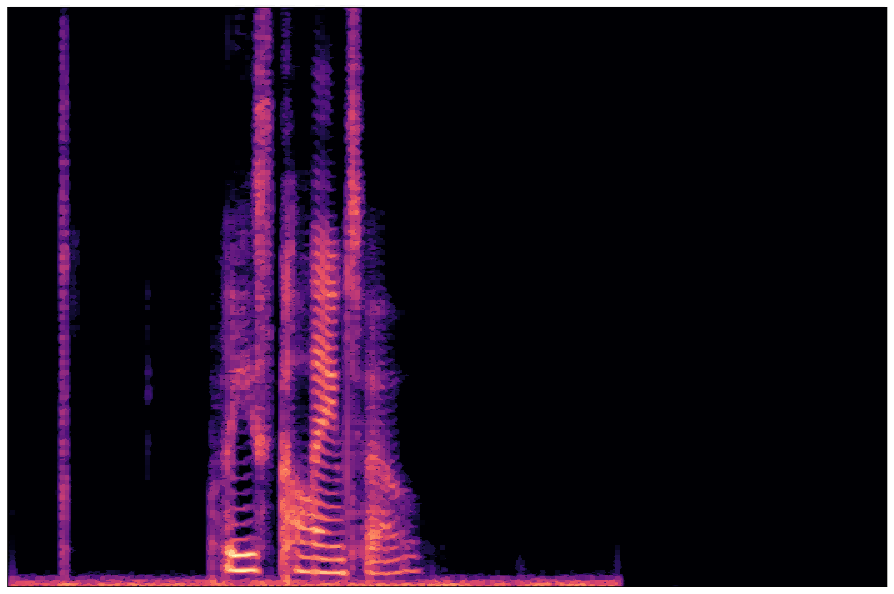}
\end{minipage}
\begin{minipage}{0.23\textwidth}
    \centering
    \includegraphics[width=\textwidth]{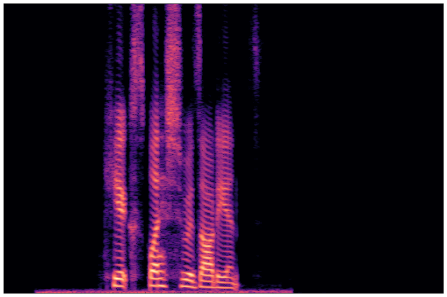} 
\end{minipage}
\caption{Separation of $\sigma(t)$ proportional guidance.}
\label{fig25}
\end{figure}

\begin{figure}[H]
\centering
\begin{minipage}{0.23\textwidth}
    \centering
    \includegraphics[width=\textwidth]{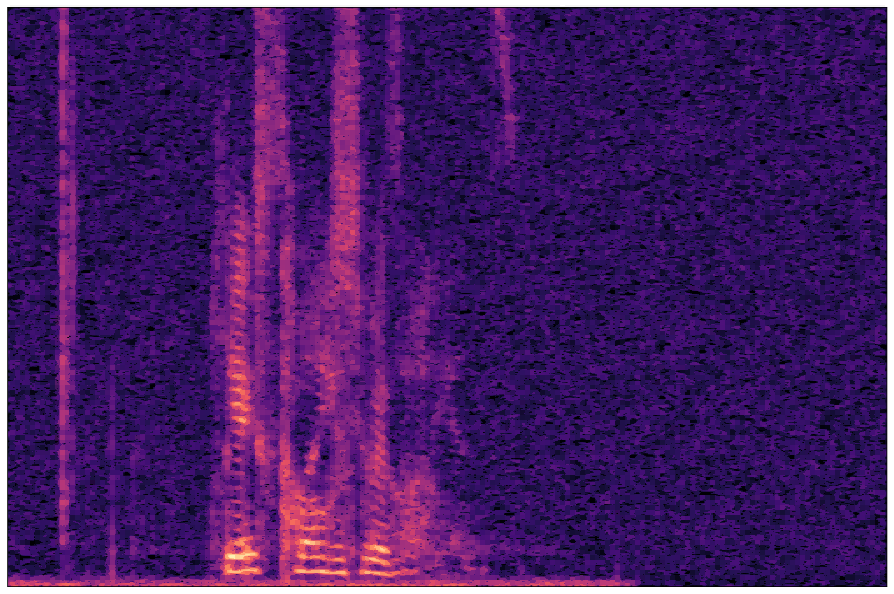}
\end{minipage}
\begin{minipage}{0.23\textwidth}
    \centering
    \includegraphics[width=\textwidth]{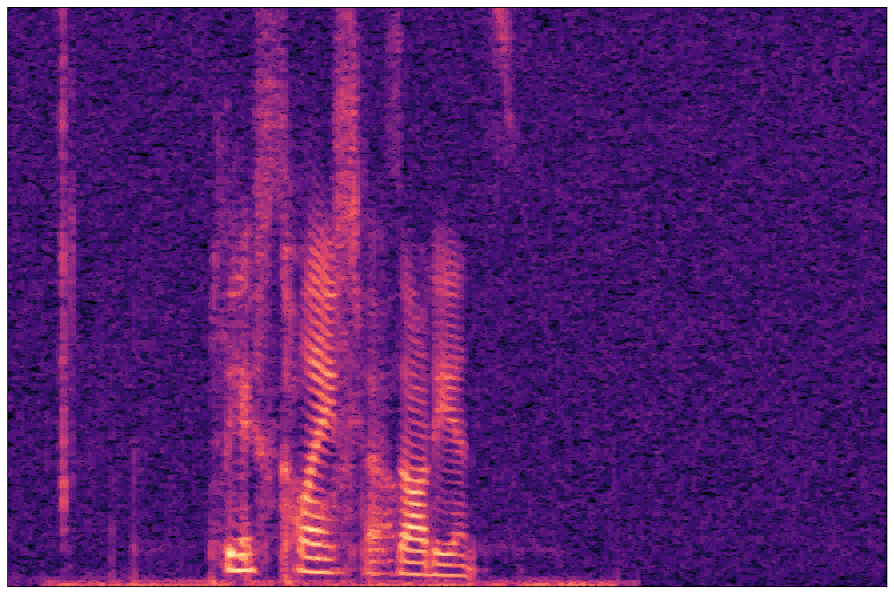} 
\end{minipage}
\caption{Separation of Analytic sampling.}
\label{fig26}
\end{figure}

\begin{figure}[H]
\centering
\begin{minipage}{0.23\textwidth}
    \centering
    \includegraphics[width=\textwidth]{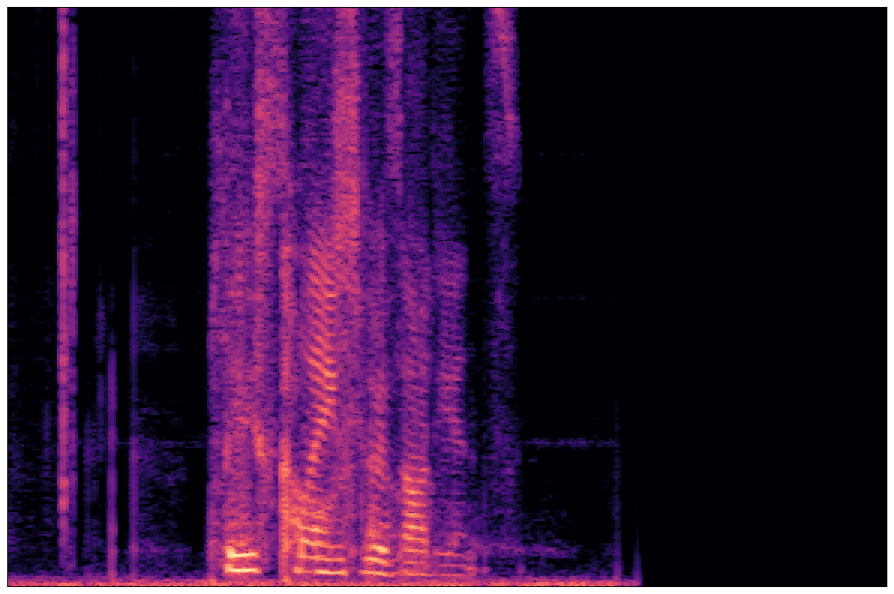}
\end{minipage}
\begin{minipage}{0.23\textwidth}
    \centering
    \includegraphics[width=\textwidth]{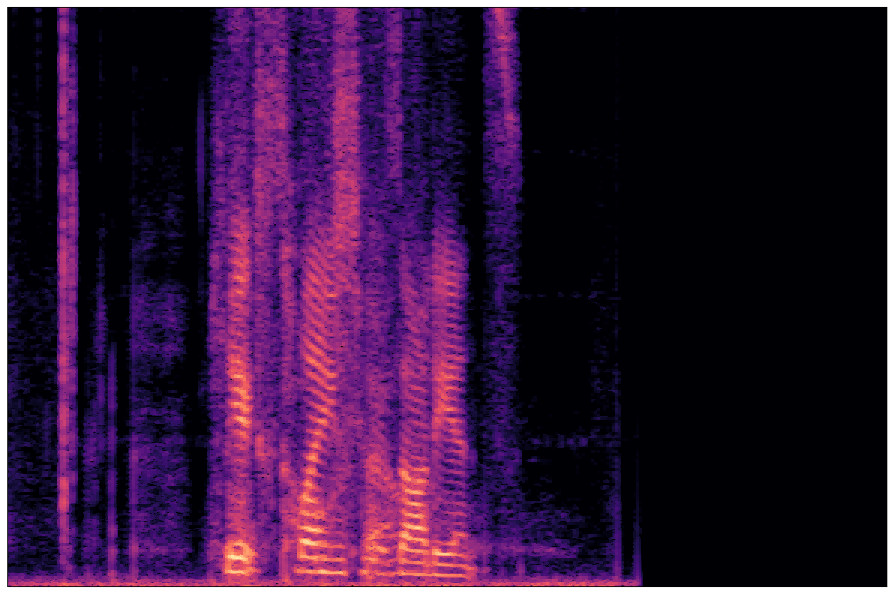} 
\end{minipage}
\caption{Separation of Conv-TasNet.}
\label{fig27}
\end{figure}

\begin{figure}[H]
\centering
\begin{minipage}{0.23\textwidth}
    \centering
    \includegraphics[width=\textwidth]{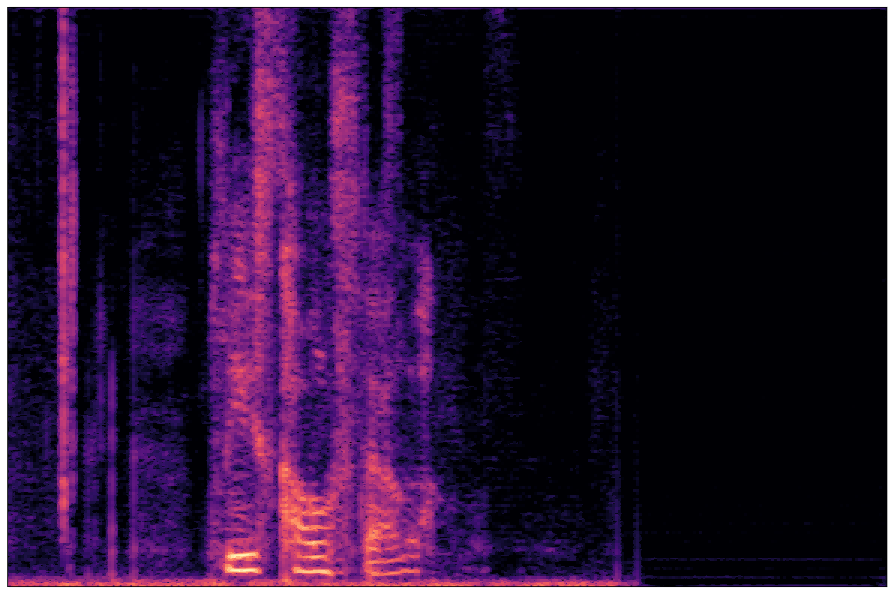}
\end{minipage}
\begin{minipage}{0.23\textwidth}
    \centering
    \includegraphics[width=\textwidth]{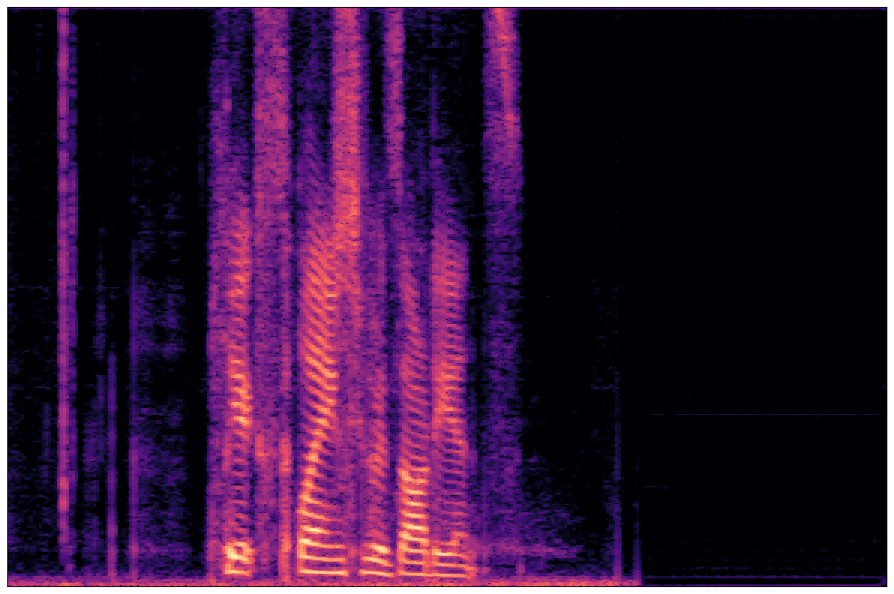} 
\end{minipage}
\caption{Separation of TF-Locoformer.}
\label{fig28}
\end{figure}


\end{document}